\documentclass[10pt,preprint]{aastex}

\newcommand {\Ha}     {H$\alpha$}    
\newcommand {\HH}     {H$_2$}        
\newcommand {\HI}     {\ion{H}{1}}   
\newcommand {\HII}    {\ion{H}{2}}   
\newcommand {\OVI}    {\ion{O}{6}}   
\newcommand {\OI}     {\ion{O}{1}}   
\newcommand {\CIII}   {\ion{C}{3}}   
\newcommand {\CII}    {\ion{C}{2}}   
\newcommand {\SiII}   {\ion{Si}{2}}  
\newcommand {\PII}    {\ion{P}{2}}   
\newcommand {\FeIII}  {\ion{Fe}{3}}  
\newcommand {\FeII}   {\ion{Fe}{2}}  
\newcommand {\NV}     {\ion{N}{5}}   
\newcommand {\CIV}    {\ion{C}{4}}   
\newcommand {\CaII}   {\ion{Ca}{2}}  

\newcommand {\kms}      {km~s$^{-1}$}

\newcommand\sk[2]{Sk\,{$-#1{^\circ}#2$}}
\newcommand\tnc{\,\tablenotemark{c}}
\newcommand\tnd{\,\tablenotemark{d}}

\begin{document}

\title{An Atlas of {\it FUSE} Sight Lines Toward the Magellanic Clouds
\footnote{This work contains data obtained for the Guaranteed Time Team by the NASA-CNES-CSA {\it FUSE} mission operated by the Johns Hopkins University.  Financial support has been provided by NASA contract NAS5-32985.}}

\author{Charles W. Danforth\altaffilmark{2}, J. Christopher Howk\altaffilmark{2}, Alex W. Fullerton\altaffilmark{2,}\altaffilmark{3}, William P. Blair\altaffilmark{2}, Kenneth R. Sembach\altaffilmark{2,}\altaffilmark{4}} 
\altaffiltext{2}{Department of Physics and Astronomy, The Johns Hopkins University, 3400 N. Charles Street,  Baltimore, MD 21218; danforth@pha.jhu.edu} 
\altaffiltext{3}{Department of Physics and Astronomy, University of Victoria, P. O. Box 3055, Victoria, BC V8W3P6, Canada}
\altaffiltext{4}{Space Telescope Science Institute, 3700 San Martin Dr., Baltimore, MD 21218}

\begin{abstract}

We present an atlas of 57 Large Magellanic Cloud (LMC) and 37 Small Magellanic Cloud (SMC) observations obtained with the {\it Far Ultraviolet Spectroscopic Explorer} satellite.  The atlas highlights twelve interstellar absorption line transitions at a resolution of $\sim$15 \kms.  These transitions cover a broad range of temperatures, ionization states, and abundances.  The species included are \OVI , which probes hot (T$\sim3\times10^5$ K) ionized gas; \CIII\ and \FeIII, which probe warm (T$\sim10^4$ K) ionized gas; \SiII, \PII, \CII, \FeII, and \OI, warm neutral gas; and six different molecular hydrogen transitions, which trace cold (T$\leq500$ K) gas.  We include Schmidt \Ha\ CCD images of the region surrounding each sight line showing the morphology of warm ionized gas in the vicinity, along with continuum images near each {\it FUSE} aperture position.

We present several initial scientific results derived from this dataset on the interstellar medium of the Magellanic Clouds and Galactic halo.  \OVI\ absorption at Magellanic Cloud velocities appears along nearly all sight lines, regardless of optical emission-line morphology.   The velocity field of LMC disk material is probed using \PII\ $\lambda$1152.8 absorption and is seen to be consistent with recent \HI\ results.  While the velocity structure of the SMC is complex, two absorption features are clearly separated in the SMC data--a strong absorption complex between $+$100 and $+$130 \kms, and a weaker feature near $+$180 \kms.  The velocity separation between these complexes varies with position, being greater on average in the southwest portion of the SMC.  A lower-velocity absorption component seen the nine sight lines toward the bright \HII\ region N66 in the SMC may be the result of an outflow or an old SNR within this nebular complex.  Absorption in \FeII\ and \OI\ at $\sim$+60 \kms\ and $\sim$+120 \kms\ appear along many LMC sight lines.  They are attributed, respectively, to an intermediate velocity cloud and a high velocity cloud in the Milky Way halo.  Both features are dramatically stronger toward the eastern half of the LMC and are not correlated with each other or with LMC \Ha\ morphology.  The lower velocity of the SMC and broader absorption lines complicate the detection of intermediate and high velocity Galactic absorption along SMC sight lines.  

\end{abstract}


\section{Introduction}

The interstellar medium (ISM) of a galaxy can be studied in many different ways. Broad-band emission-line imaging is rich in morphological detail, but contains little kinematic information. Furthermore, it is generally sensitive to material of intermediate temperature and density and so cannot address the whole range of expected conditions within the ISM.  Emission-line spectroscopy at sufficiently high spectral resolution can provide kinematic information, but again only for a limited selection of transitions.

Absorption-line spectroscopy is a powerful tool for analyzing the ISM and has several inherent advantages over either imaging or emission-line spectroscopy.  Absorption-line spectroscopy complements emission-line measurements; it provides sensitivity to very low column densities and much broader range of diagnostics for gas at differing temperatures and densities.  The far ultraviolet (FUV; $\lambda < 2000$ \AA) is particularly rich in diagnostic lines, and the number of available lines increases dramatically with decreasing wavelength.  Transitions of \OVI, \NV, and \CIV\ probe hot ($1-3 ~ \times ~10^5$ K) gas present in shocks and transition layers between hot and cold components of the ISM.  Ground-state molecular hydrogen transitions (found only below 1108 \AA) provide information about cold (T=50-500 K), dense gas.  Low-ionization O, C, and N transitions probe the warm (T=$10^2-10^4$ K) neutral and photoionized ISM.  Less-abundant species such as Fe, Si, Ar, P, S, and Cl yield information about the elemental abundances and chemistry of the ISM.

Extinction in the FUV and the low efficiency of UV telescopes have been the main constraints on FUV observatories.  {\it Copernicus} was limited to observing a few relatively bright stars at distances of less than a kiloparsec \citep{Morton73,Rogerson73}.  With the {\it International Ultraviolet Explorer} ({\it IUE}), long exposures were used to observe the brightest extragalactic stars, such as O giants in the Large Magellanic Cloud (LMC) \citep{SavagedeBoer79,SavagedeBoer81}.  The {\it Hopkins Ultraviolet Telescope (HUT)} \citep{HUTref} featured improved sensitivity but at a lower spectral resolution.  {\it ORFEUS} \citep{Orfeusref} had higher resolution, but poor sensitivity.  Both {\it ORFEUS} and {\it HUT} were shuttle-borne payloads and only observed for a few weeks each.  The {\it Far Ultraviolet Spectroscopic Explorer} ({\it FUSE}) mission represents a significant improvement in spectral resolution, collecting area, and sensitivity over these earlier missions covering the FUV \citep{Moos00,Sahnow00}.  {\it FUSE} is a dedicated spectroscopic observatory covering the wavelength range 905-1187 \AA\ at a resolution of R$\equiv \lambda/\Delta\lambda\approx$15,000-20,000.  The sensitivity of {\it FUSE} allows high-quality spectra of individual O and B stars in the LMC and Small Magellanic Cloud (SMC) to be obtained with modest exposure times.


The Magellanic Clouds are ideal laboratories for the study of ISM morphology, abundances, ionization, and kinematics.  Both galaxies are rich in interstellar material with hundreds of \HII\ regions, bubbles, supershells, supernova remnants (SNRs), and areas of diffuse emission.  Given their proximity and the fact that each galaxy spans several degrees across the sky, they can be studied globally and with great attention to the details of individual sub-structures.  Since the LMC and SMC lie, respectively, 33$^\circ$ and 45$^\circ$ out of the Galactic plane, there is relatively little foreground extinction from our own Galaxy, allowing FUV and X-ray observations to be made efficiently.

The LMC is the nearest disk galaxy to the Milky Way.  Though it is optically irregular, \HI\ images \citep{Kim98} show the LMC to be a roughly face-on spiral disk with an inclination angle of $\sim35^\circ$ \citep{Westerlund97}.  The geometry of the SMC is a subject of continuing debate \citep{Mathewson88,Staveley-Smith97}, but it is clear that the morphology is very much different from either the Galaxy or the LMC.  The SMC is the closest example both a gas-rich dwarf galaxy and a galaxy that has undergone significant tidal interactions.  Both Clouds have star formation histories and mean abundances significantly different from each other and from the Milky Way.

While the Magellanic Clouds themselves are substantially different in morphology, metallicity, star-formation history, and many other parameters, many of the smaller scale features found in the Magellanic Clouds have direct analogues in our own Galaxy.  For many categories of objects the available Magellanic Cloud samples are larger and easier to study systematically than the Galactic samples, which may be obscured in the disk, at unknown distances, or beset by line-of-sight confusion.  With the sensitivity of modern instruments, even individual stars of types rarely observable in our own galaxy -- Wolf-Rayet (WR) stars, luminous blue variables, etc. -- are readily detectable in the relatively unreddened Magellanic Clouds.  The SNRs N132D and 0540-69.3 in the LMC and E0102$+$7219 in the SMC share similar morphology with, respectively, the Cygnus Loop, the Crab Nebula and Cas A in the Milky Way.  Larger structures such as superbubbles and supergiant shells \citep{Meaburn80,Staveley-Smith97,Kim99} are also obvious in the Clouds.  These features are likely similar to Milky Way structures such as the Scutum supershell \citep{Callaway00}, the W4 Chimney \citep{Reynolds01}, and the Aquila supershell \citep{Maciejewski96}.


In the first two years of operation, {\it FUSE} has observed over one hundred hot stars of various types in the Magellanic Clouds.  These observations provide an incredibly rich dataset for studying stellar and interstellar physics.  Each observation can be used to study the photosphere of the star itself and the intervening material which imposes its fingerprint on this spectrum.  Since the systemic velocity of each Magellanic Cloud is substantially offset from the Milky Way (typically $+$200--300 \kms\ for the LMC and $\sim$+150 \kms\ for the SMC), one can simultaneaously study both the Galactic material and that of the host Magellanic Cloud in each spectrum.

In this paper we provide an overview of the large database of {\it FUSE} Magellanic Cloud observations in the form of an atlas of ISM FUV absorption line spectra and \Ha\ emission-line images for 94 sight lines.  The spectra and images in this atlas are useful for studying the large-scale characteristics of the gaseous component of the Magellanic Clouds and the Milky Way halo in this general direction.  In \S2 we discuss target selection, the optical imaging, and spectral line selection for the atlas.  In \S3 we discuss the data reduction and calibration steps used to create the atlas.  While detailed analyses will be presented in later papers, \S4 discusses a few initial findings from this large dataset.

\section{Observations}
\subsection{Target Selection}
Of the more than one hundred stellar targets observed in the Magellanic Clouds by {\it FUSE}, we have selected those observed under {\it FUSE} Team Guaranteed Time programs as well as selected guest investigator data that are now in the public domain.  The program names and principal investigators are listed in Table~1.  We then culled datasets with very low signal-to-noise and those that did not have data in all four spectrograph channels.  The resulting sample consists of 57 targets in the LMC and 37 in the SMC.  All of the stars are classified as spectral type WR, O, or early B (B2 and earlier).  Hot stars have strong UV continua and tend to have strong winds and high rotation velocities yielding broader stellar spectral features.  The relatively narrow ISM features are then easier to identify against the spectra of early-type stars and the stellar continua are easier to characterize.  

We list the 94 targets in Table~2 along with their stellar parameters.  For LMC targets we adopt the naming and ordering convention of \citet{Sanduleak70}: whole degrees of declination followed by increasing right ascension.  For the SMC targets, we use the naming convention of \citet{AzV}: sorted with increasing RA.  In the cases where Sk or AV numbers do not exist, a name from a different nomenclature has been supplied.  Alternate names appearing in the literature are also provided in many cases.

To better understand the spatial distribution of the sight lines and their relationship to optical emission-line morphology, we have overlayed their positions on \Ha\ images of the two Clouds.  Figure~1 shows the LMC targets overlayed on a continuum-subtracted \Ha\ image of the LMC \citep{SHASSAref}.  Figure~2 shows the same for the SMC \citep{SHASSAref}.  Some densely-clustered observations such as those in the heart of 30 Dor, N51D, or NGC\,346 may appear as a single star in the figures.

\subsection{Optical Imagery}
To help characterize the morphology and interstellar environment in which each {\it FUSE} target is located, we obtained wide-field, emission-line CCD images with the Curtis Schmidt Telescope at CTIO in 1999 November.  Fields containing most of the {\it FUSE} sight lines presented here were imaged with a 2048$\times$2048 pixel CCD at $\sim$3\arcsec\ resolution through \Ha\ ($\lambda_c$=6563\AA, $\Delta\lambda$=30\AA) and continuum ($\lambda_c$=6850\AA, $\Delta\lambda$=95\AA) filters.  A more comprehensive survey using the same instrumentation and an expanded filter set is being undertaken by \citet{Smith99}.  Images were reduced, aligned, and coadded using standard IRAF\footnote{IRAF is distributed by the National Optical Astronomy Observatories, operated by the Association of Universities for Research in Astronomy, Inc., under cooperative agreement with the National Science Foundation.} procedures.  

Typically, three exposures of five minutes each were coadded in the \Ha\ band; two exposures of one minute each were used for the continuum images.  After aligning and coadding separate exposures, a series of 512$\times$512 pixel sub-images centered on each {\it FUSE} target or group of targets were extracted.  Unfortunately, some of the more recent {\it FUSE} targets were not imaged during our observing run in 1999.  Astrometric solutions were applied to each subimage using the plate scale and sky and pixel coordinates of the {\it FUSE} target in each image.  These solutions are approximate and are intended only as a reference.

\subsection{Spectral Data}
Descriptions of the design and performance of the {\it FUSE} spectrographs are given by \citet{Moos00} and \citet{Sahnow00}.  All of the observations included in the atlas were obtained in time-tag mode and most were observed through the $30\arcsec\times30\arcsec$ (LWRS) apertures (the exceptions are \sk{67}{106} (A11101), \sk{67}{107} (A11102), \sk{68}{75} (A04905), and AV332 (P10304), which were observed in the $4\arcsec\times20\arcsec$ (MDRS) aperture).  Most observations were broken into multiple exposures taken over consecutive orbital viewing periods.  The photon lists from these exposures were concatenated before being processed by the standard {\it FUSE} calibration pipeline software, {\sc calfuse} (v1.8.7).  

A detailed description of these processing steps is provided in the {\it FUSE} Data Handbook{\footnote{Available at {\tt http://fuse.pha.jhu.edu/analysis/dhbook.html}}}.  In brief, the {\sc calfuse} software computes the shifts in the detected position of a photon required to correct for: (a) the motion of the satellite; (b) thermally-induced motions of the diffraction gratings; (c) small, thermally-induced drifts in the read-out electronics of the detector; and (d) fixed geometric distortions in the detector.  Application of these shifts produces a two-dimensional, distortion-corrected image of a detector segment, from which a small, uniform background is subtracted.  One-dimensional spectra are extracted for the LiF and SiC channels recorded on each of the four detector segments.  

After correcting for detector dead time (which is very small for count rates typical of targets in the Magellanic Clouds), the most recent wavelength and effective-area calibrations are applied to convert the detector count rate in pixel space to flux units as a function of wavelength.  The detector pixel scale ($\sim$6.7 m\AA\ for LiF, $\sim$6.2 m\AA\ for SiC) over-samples the instrumental resolution of $\sim$15--20 \kms.  The spectra were rebinned by three pixels---roughly half a resolution element for this atlas. 

The {\it FUSE} observational parameters of all 94 targets are listed in Table~3.  Column~2 lists the {\it FUSE} target identifier (program and target number).  The third column lists the observation numbers used in this atlas, and the fourth column lists the total number of coadded exposures used.  The fifth column lists the total exposure time for the target, and the sixth column lists the date of the earliest exposure.

Accurate absolute wavelength calibration to better than 10--15 \kms\ for {\it FUSE} data is very difficult.  However, the relative wavelength solution for {\it FUSE} is well-characterized.  The {\sc calfuse} pipeline applies the wavelength solution and corrects for spacecraft orbital motion and Earth orbital motion.  This nominally leaves the velocity in a heliocentric rest frame, but a sign error in the geocentric to heliocentric correction in {\sc calfuse} v1.8.7 and earlier leaves the solution incorrect by twice the geocentric-to-heliocentric velocity.  In processing the data for this paper, we corrected this error manually.  Later versions of {\sc calfuse} correct this problem.  We have applied a heliocentric-to-LSR correction assuming a solar motion of $+$20 \kms\ in the direction l=56$^\circ$, b=22$^\circ$.

Finally, a linear offset was made to the velocity scale of each observation to bring the data into the LSR frame.  This offset was determined by comparing {\it Space Telescope Imaging Spectrograph (STIS)} \FeII\ $\lambda$1608 absorption lines (which were calibrated with an onboard lamp) to four \FeII\ transitions in the {\it FUSE} range observed in the LiF1 channel.  Using two LMC targets (\sk{67}{104} and \sk{71}{45}) and eight SMC targets (AV15, AV47, AV69, AV75, AV83, AV95, AV327, and AV423) we derive an offset of $-47.4\pm5.6$ \kms\ from the given {\it FUSE} absolute wavelength solution.  The spectra for all targets in the atlas were shifted by this amount.  

Due to thermal control issues \citep{Sahnow00}, all four {\it FUSE} channels are not always in perfect alignment.  Most of our data is taken from the LiF1 channels.  The exception is the \CIII\ $\lambda$977.020 transition which was observed with SiC2 and hence may be offset in velocity from the LiF1 data.  To correct the channel alignment, we have cross-correlated the region around \SiII\ $\lambda$1020.699 in the LiF1 and SiC2 channels for each observation.  The resulting shifts were applied to the \CIII\ $\lambda$977.020 segments shown in the atlas figures.

A check was performed on our velocity calibration routine by looking at the velocity of Galactic absorption seen in the sight lines.  The \PII\ $\lambda$1152 shows the narrowest absorption profiles of any in our data.  For the LMC sample, the Galactic absorption appears at $-8.1\pm4.5$ \kms\ (LSR).  This velocity dispersion is considerably smaller than the {\it FUSE} resolution.  {\it STIS} measurements of two sight lines in the LMC show \FeII\ absorption between $-5$ and $-10$ \kms.   Thus, we estimate an absolute velocity uncertainty of $\pm10$ \kms\ for the data as presented here.

While many interesting ISM lines can be found in a typical target spectrum, we have chosen nine ions to highlight in this study;  these are summarized in Table~4.  They represent some of the more useful transitions for ISM studies covering a broad range of temperatures, ionization states, and abundances.  In addition, we have included three molecular hydrogen lines arising from different rotational levels.  Not only does this provide valuable information about the kinematics of cold, dense gas, but it lets the molecular hydrogen `contamination' of ionic transitions be assessed in a straightforward manner.  \citet{Tumlinson01} present a comprehensive {\it FUSE} survey of interstellar \HH\ in the Magellanic Clouds.  

\section{The Atlas}
In Figures~3--59 we present the atlas of {\it FUSE} sight lines in the Large Magellanic Cloud.  Figures~60--96 show the atlas of SMC sight lines.  Since each atlas page presents a large quantity of information, we will describe in detail what is shown in these figures.  Information about each target is displayed in the upper right corner of each figure: sky coordinates, spectral type, V magnitude and E(B$-$V).  Also shown is various {\it FUSE}-related information: {\it FUSE} archive identifier, aperture (either LWRS for the large aperture or MDRS for the medium aperture), number of exposures coadded, total exposure time in seconds, and the date of the first exposure.  Since many of our targets have many common names in the literature, we include alternate names below the {\it FUSE} parameters.

The images in the upper left-hand corners of Figures~3--96 are taken from the CTIO Schmidt \Ha-band images discussed in \S2.2.  Each image is $\sim$19\arcmin\ on a side, corresponding to physical dimensions of $\sim$280 pc ($\sim$335 pc) at the distance of the LMC (SMC).  We assume a distance of 50 kpc for the LMC and 59 kpc for the SMC.  These images have not been continuum-subtracted; thus they trace the morphology of the warm (T$\sim10^4$K) \Ha-emitting material as well as the positions of stars through their optical continuum emission.  The {\it FUSE} target is marked with a 1\arcmin\ box (see below).  Other {\it FUSE} pointings included in this paper that fall in the same field are marked with stars.  Each image is displayed logarithmically and display levels are automatically adjusted to highlight the nebular morphology in the immediate vicinity of the {\it FUSE} target.  

The two smaller images below the summary information show the central arcminute around the {\it FUSE} target in detail in the continuum and \Ha-bands.  Superimposed on the boxes are {\it FUSE} apertures showing the spacecraft roll angle and aperture location during the observations.  The ticks on either side of the aperture show the dispersion direction for the observation.  These close-up images are useful for assessing whether there are other bright stars or obvious structures within the {\it FUSE} aperture that may contribute to the FUV flux.  Multiple stars in the spectrograph aperture can lead to broadened and shifted absorption features.  This is an important issue in the {\it FUSE} data.  Most of the {\it FUSE} observations were made through the 30\arcsec$\times$30\arcsec\ LWRS aperture.  In most cases a given {\it FUSE} pointing will capture only one hot star bright in the FUV, but for some pointings in the Magellanic Clouds, multiple stars may appear in the aperture.  For example, in the same aperture as \sk{69}{104} (Figure~42) a second star---the O4 giant LH41-32~---can be seen about 15\arcsec\ to the south-east.  LH41-32 and \sk{69}{104} are both early-type stars and produce similar FUV continua.  However, being $\sim$15\arcsec\ apart along the dispersion direction of the aperture, they produce diluted absorption features $\sim$60 \kms\ apart.

More dramatic examples of stellar contamination can be found in the cores of hot, dense clusters.  In 30 Doradus in the LMC, dozens of early-type stars contribute to the FUV flux and make the spectra of Melnick 42 (Figure~49) and \sk{69}{243} (Figure~50) essentially indistinguishable.  When observed through the LWRS aperture the result is a set of very broad, washed out absorption features resulting from the distribution of FUV flux within the aperture.  Any ISM analysis using these crowded fields must take the severe blending of narrow ISM lines into account.  The dense SMC cluster NGC\,346 hosts four FUSE targets within an arcminute of each other, but unlike the 30 Dor sightlines, these targets were observed with the MDRS aperture to minimize stellar contamination.

The bottom two thirds of each page in Figures~3--96 are occupied by twelve spectral segments featuring the transitions listed in Table~4.  Each segment is presented in velocity space, covering $-$400 to $+$600 \kms\ on the velocity scale appropriate for the transition noted in the bottom left of each panel.  All velocities are in the LSR frame within the uncertainties discussed in \S2.3.  The spectra are displayed in unnormalized flux units of $\rm10^{-13}~erg~cm^{-2}~s^{-1}~\AA^{-1}$.

Dashed vertical index lines mark the average systemic velocities for the Galaxy and Magellanic Clouds (v$\rm_{LSR}$=+264 \kms\ for the LMC \citep{Kim98} and v$\rm_{LSR}$=+165 \kms\ for the SMC \citep{Staveley-Smith97}).  LMC and SMC absorption features may appear over a velocity range of $\pm$50 \kms\ or more;  hence the actual components in a given sight line may differ from the average systemic velocities.

Above the spectra in each panel is a series of small annotations.  These mark the expected locations of other absorption features which may, but do not necessarily, appear in the data.  Galactic (v=0 \kms) components are labeled in normal type face while Magellanic (v=+264 or $+$165 \kms) components are in an italic type face positioned slightly lower.  The same velocities are used for these offsets as are used for the primary transitions, as discussed above, and the same caveat about velocity ranges applies.  The notation `R4', `P3', etc., is a shorthand description for molecular hydrogen transitions.  For example, the `R4' transition seen in the \OVI\ panel is shorthand for the (6-0) R(4) ro-vibrational transition of \HH\ at 1032.356\AA\ from J=4 to J=5 and v=0 to v=6.  The `P3' transition in the same panel represents the (6-0) P(3) 1031.191\AA\ transition from J=3 to J=2 and v=0 to v=6.  Wavelengths and f-values for these secondary transitions are given in Table~5.  

The first transition covered in Figures~3--96 (upper left panel) is \OVI\ $\lambda$1031.926, an important transition for studies of hot ($\sim3\times10^5$K) gas.  {\it FUSE} is the first instrument to have the capability to study \OVI\ at high resolution for many stars more than a few kiloparsecs from the sun.  \OVI\ is a doublet but the other transition at $\lambda$1037.617 is frequently blended with several other strong absorption lines along Magellanic Cloud sightlines. 

The next two panels show \CIII\ $\lambda$977.020 and \CII\ $\lambda$1036.337.  \CIII\ is a very strong transition and is always heavily saturated.  However, with an ionization potential of 24.383 eV, \CIII\ is the next-highest ionization species common in the ISM in the {\it FUSE} range.  \CII, like \CIII, is a strong transition of an abundant element and is always strongly saturated.  With these two saturated transitions, we can probe for very low column density absorption features at high and low velocities.

The fourth panel is devoted to \SiII\ $\lambda$1020.699.  This is the strongest \SiII\ transition in the {\it FUSE} band and gives information about dust destruction in the ISM.  Next is \PII\ $\lambda$1152.818.  This is a strong transition, but of a low-abundance element;  thus it has small equivalent widths and is useful for finding velocities of the highest column densitiy Magellanic Cloud features.  \PII\ is not heavily depleted onto dust grains, and the region around 1152\AA\ is uncontaminated by other transitions.

The bottom panel on the left side shows the strong \OI\ 1039.230\AA\ transition.  \OI\ has a similar ionization potential to atomic hydrogen and should show very similar velocity structures.  The strength of \OI\ $\lambda$1039.230 makes it a useful probe of low-column features such as HVCs and IVCs.  While many observations were carried out primarily at night, geocoronal \OI\ airglow may contaminate the region around v=0 \kms\ in some spectra.  LMC \HH\ absorption may appear near v=+100 \kms\ and SMC \HH\ absorption may appear near v=0 \kms\ in this panel.

The top three panels on the right side of Figures~3--96 are devoted to iron ions.  \FeIII\ $\lambda$1122.524 is one of the only clean examples of a doubly-ionized species in the FUSE band.  It is a triplet of overlapping lines arising from the ground state of doubly-ionized iron.  Galactic \FeIII\ absorption can be contaminated by \FeII\ $\lambda$1121.975.  LMC \FeIII\ is usually clean, but can be compromised by a stellar P-Cygni profile from \ion{P}{5} and \ion{Si}{4} for some targets.

\FeII\ $\lambda$1144.938 is a strong iron transition uncontaminated by other lines.  Along with \OI, it tracks low-column gas well.  \FeII\ $\lambda$1125.448 is a weaker transition of the same ion and shows sharper absorption features for the higher-column components.

The bottom three panels are devoted to molecular hydrogen transitions.  While the main focus of this work is the warm and hot gas, these transitions help calibrate the \HH\ contamination present in the rest of the spectra; see \citet{Tumlinson01}.  The first \HH\ panel focuses on the (0-0) R(1) $\lambda$1108.634 transition of \HH, but it also contains the overlapping (0-0) R(0) transition at $\lambda$1108.128.  Both transitions have similar oscillator strengths and give insight into the dynamics of the cold molecular gas.  The (1-0) P(5) transition at $\lambda$1109.313 probes a more excited state of \HH, but absorption in this line is usually very weak or absent.

The second \HH\ panel also contains two overlapping transitions: (4-0) R(2) $\lambda$1051.497 and (4-0) P(1) $\lambda$1051.031.  The former is a strong transition that probes slightly more excited gas than the two transitions in the panel above.  The latter transition is weak and probes the same gas as (0-0) R(1) $\lambda$1108.634.

The final panel shows another two partially blended \HH\ transitions: (5-0) R(4) at $\lambda$1044.546 and (5-0) P(3) $\lambda$1043.498.  These transitions arise from the same states as those that can contaminate the \OVI\ $\lambda$1031.926 line and are thus of special interest in the analysis of \OVI.  These lines represent a higher excitation level of cold \HH.

We will use the LMC target \sk{65}{22} (Figure~4) as an example of how one can use the information in the atlas.  \sk{65}{22} is an interesting target with excellent signal to noise observations and an interesting mix of ISM features.  In the upper left-hand spectral panel, we see definite signs of \OVI\ absorption at $\sim$0 \kms\ arising in the Galaxy and at $\sim$+250 \kms\ from the LMC.  The sharp absorption feature at $\sim$+120 \kms\ labeled `R4' is Galactic \HH\ arising from the J=4 level.  This is in agreement with the information given in the bottom right-hand panel where we see that there is moderate \HH\ (5-0) R(4) absorption at Galactic velocities and essentially none at LMC velocities.  Returning to the \OVI\ panel, we see that the R(4) at LMC velocities (seen in this panel at $\sim$+380 \kms), is very weak, in good agreement with the bottom right-hand panel.  The other \HH\ transition labeled in the \OVI\ panel is the (6-0) P(3) transition.  Again, a similar transition is covered in the lower right panel of this figure.  It is easy to see that there is a strong absorption from \HH\ (5-0) P(3) $\lambda$1043.498 at Galactic velocities and somewhat weaker absorption at LMC velocities.  By comparing the two wavelength regions, we can calibrate how much, and at what velocities, the \HH\ absorption affects the \OVI\ and other lines.  

This approach is useful in other cases as well.  Using the same figure, we see several absorption features in the \OI\ panel at $\sim$+120 \kms\ coincident with an LMC \HH\ R(2) feature.  By checking the R(2) panel, we see that there does appear to be some LMC \HH\ present.  But a feature is also present at v$\sim+$120 \kms\ in \FeII\ $\lambda$1145, which is unaffected by \HH, implying that at least some of the apparent intermediate-velocity absorption comes from intermediate velocity material, not LMC \HH.  In this case, the absorption is due to a high velocity cloud (HVC) in the Galactic halo.  This feature is seen frequently in strongly resonant transitions (\OI\ $\lambda$1039 and \FeII\ $\lambda$1145) at $\sim+$120 \kms\ and is discussed more thoroughly in \S4.2.

\section{Discussion}
Selected spectral lines observed by {\it FUSE} reveal a variety of velocity components and physical conditions in the gas along the Magellanic Cloud sight lines.  We can gain an understanding of the ISM structure of the Magellanic Clouds by looking for variations between sight lines and variations within a sight line over different temperature regimes.  We present here a few initial findings.

\subsection{The Ubiquity of Magellanic \OVI}
One of the most striking initial findings from the atlas is that \OVI\ absorption associated with the Magellanic Clouds is found along practically all sight lines in both Clouds.  Quantitative data on column densities and velocity are difficult to obtain in some cases; the same hot stars that produce FUV flux also produce strong winds that often make the stellar continua difficult to determine near 1032 \AA, particularly for stars later than spectral type O7.  Nevertheless, inspecting Figures~3--96 it is clear that most, if not all, sight lines contain substantial \OVI\ absorption at Magellanic Cloud velocities.

\citet{Howk01} measured the \OVI\ column depth to twelve LMC stars for which a stellar continuum could be reliably placed.  They find that the \OVI\ column depth has little correlation with morphological features in either \Ha\ or X-rays.  For example: \sk{67}{20} (Figure~14) is a field star far from any enhanced \Ha\ or X-ray emission, while \sk{66}{100} (Figure~7) lies interior to the \Ha\ and X-ray bright supergiant shell LMC4.  Yet the column depth of LMC \OVI\ toward both stars is identical ($\rm log N($\OVI$\rm)=14.26\pm0.05~cm^{-2}$).  In the bulk of the datasets the stellar continuum near 1032 \AA\ is too ambiguous for quantitative analysis, but it is clear that \OVI\ absorption at Magellanic Cloud velocities is present in most sight lines regardless of morphological setting.

Furthermore, Howk et~al. showed that \OVI\ absorption is present over a broader range of velocities than lower ionization gas along the same sight line.  Typically, the high velocity edge of the \OVI\ absorption will coincide with the high velocity edge of \FeII\ and other cooler ion absorption.  But \OVI\ extends to lower velocities than other species.  \sk{67}{211} (Figure~30) shows this well with a broad \OVI\ absorption featured with a centroid velocity of $\sim$+240 \kms.  In the same sight line, the \FeII\ $\lambda$1125 absorption centroid is at $\sim$+270 \kms.  The uniformity in \OVI\ absorption despite differences in optical morphology and the broad, lower-than-systemic velocity absorption of \OVI\ toward the LMC suggests that there is a hot halo around the entire LMC rather than just numerous individual bubbles of hot gas within the disk \citep{deBoerSavage80}.

\citet{Hoopes01b} studied \OVI\ absorption in the SMC and found a similar ubiquity of hot gas.  But the \OVI\ column density correlates with position in the SMC, with the largest values occurring near bright star-forming \HII\ regions along the bar.  Away from star-forming regions there are smaller, but still substantial, amounts of \OVI.  Even in these regions the \OVI\ column depth is roughly equal to that in the Milky Way along the same sight line.  Unlike the LMC, SMC \OVI\ absorption occurs near the systemic velocity.  This supports the idea that the SMC also has a hot halo but increased star formation has produced substantial \OVI\ in the disk as well.

\subsection{Intermediate and High Velocity Absorption}

We examined the \FeII\ $\lambda$1145 line for absorption features in all LMC sight lines, using the \OI\ as a secondary comparison line.  This \FeII\ transition, which has a large oscillator strength, shows low column density absorption and is uncontaminated by other absorption lines.  Some stars were discarded from the analysis due to ambiguous continua and/or low signal to noise in the 1145\AA\ region.  We fit \FeII\ profiles for the remaining sight lines with multiple Gaussian components after fitting a low-order polynomial to the local stellar continuum.  For each absorbing complex along a sight line, our fits yielded the central velocity and equivalent width.

Figure~97 shows a histogram of all velocity components measured in this way.  Low velocity Galactic absorption lies in the velocity range $-$30 to $+$20 \kms\ with a dispersion of 6 \kms, while LMC absorption appears from $+$185 to $+$340 \kms\ with a much larger dispersion of 29 \kms.  Two other distinct populations appear as well.  Between $+$20 and $+$75 \kms\ we see an intermediate velocity component with a dispersion of 10 \kms.  This corresponds well with the ``60 \kms'' component observed in previous studies \citep{SavagedeBoer79,SavagedeBoer81,Wayte90} that has been attributed to an intermediate velocity cloud (IVC) in the Galactic halo.

A higher-velocity component is readily apparent in many sight lines between $+$75 and $+$135 \kms\ (with a dispersion of 14 \kms).  This has also been widely observed and corresponds to Savage \& de Boer's ``130 \kms'' component (Savage \& de Boer use heliocentric velocities which differ from the LSR frame in this direction by $\sim$15 \kms.)  It is unlikely that this absorption arises in tidally stripped material from the LMC since the velocity and velocity dispersion of the absorption features is so low.  Savage \& de Boer instead attribute this feature to a high velocity cloud (HVC) present between the Galaxy and the LMC. 

Our measurements show that neither the IVC nor the HVC absorption correlates with LMC morphology.  Figure~98 (left panel) shows the distribution of the IVC feature across the disk of the LMC;  HVC absorption is shown in the right panel.  Crosses represent all sight lines used in this analysis.  Circles surrounding the crosses indicate IVC or HVC absorption with the radius of the circle proportional to the measured equivalent width of the line in \FeII\ $\lambda$1145.  Sight lines with concentric circles indicate multiple fit components within the velocity range.

IVC absorption appears fairly evenly across the eastern half of the LMC while the western half shows only occasional absorption from this component.  Evidently, we are seeing the edge of a cloud in the Galactic halo.  HVC absorption is a bit more patchy but appears strongest in front of the northeastern part of the LMC.  The strength of HVC absorption shows more variation than the IVC component and similarly does not correlate with the morphology of the LMC itself.  While these two spatial distributions appear similar, there is no correlation in equivalent width between IVC and HVC absorption features.  Furthermore they appear as distinct distributions in Figure~97 with distinct dispersions showing that they are, in fact, independent systems.  

Savage \& de Boer noted the presence of a ``170 \kms'' feature which occasionally appeared in their {\it IUE} data.  We also see corresponding absorption toward the LMC between $+$135 and $+$185 \kms.  This ``very high velocity cloud'' appears in the spectra of 13 stars (\sk{66}{18}, \sk{66}{51}, \sk{67}{20}, \sk{67}{69}, \sk{67}{166}, \sk{67}{167}, \sk{67}{211}, BI272, \sk{68}{80}, \sk{68}{82}, \sk{68}{135}, \sk{69}{59} and \sk{69}{246}) although some of the detections are marginal.  \citet{Wayte90} interprets this component as gas within the LMC;  we draw the same conclusion from more data points.  The ``170 \kms'' absortion is frequently associated with regions of LMC nebulosity and the velocity is only $\sim$60 \kms\ different from that of LMC material.  This suggests a Magellanic origin for this material, perhaps a patchy outflow from \HII\ regions.

IVC and HVC absorption may also exist toward the SMC but they are not obvious in the data.  The velocity structure of the SMC is complex with several distinct components spread over nearly 100 \kms.  Furthermore, the systemic velocity of the SMC is lower than the LMC and it would be difficult to distinguish HVC absorption at $\sim$+100 \kms\ from the wing of a saturated SMC component at similar velocities.

\subsection{The Velocity Structure of the SMC}
It is known that the SMC has a very complicated velocity structure in \HI\ \citep{Staveley-Smith97,Stanimirovic99}.  The two massive components making up the SMC are seen near v$\rm_{LSR}$=$+$125 and $+$156 \kms\ \citep{Hindman67,McGeeNewton81,Mathewson84,Mathewson88,Staveley-Smith97}.  The stellar population and ionized gas distribution has been seen to follow the same pattern \citep{Mathewson86}.  Similarly, \CaII\ line splitting has been observed toward many SMC stars \citep{Wayte90}.  \citet{MuraiFujimoto80} suggest that these different velocity components arose when the SMC was torn in half during a close passage with the LMC $2\times10^8$ years ago.  They posit that tidal disruption from the Milky Way is continuing to draw the various pieces apart.  The opposing view, first proposed by \citet{Hindman67} and more recently revised by \citet{Staveley-Smith97}, argues that the velocity components are caused by expanding shells within the ISM of the SMC.

Two additional, weaker velocity components are sometimes seen in \HI\ \citep{McGeeNewton81,Mathewson88} at $+$105 and $+$182 \kms.  The $+$105 \kms\ component is strongest in the central and southwestern bar.  Occasionally, it is as strong as the two main components.  The $+$182 \kms\ component is weaker than the other three and is concentrated in the bar.

The most obvious feature seen in the SMC {\it FUSE} data is that many of the low-ionization transitions show doubled absorption profiles at SMC velocities.  There is always strong SMC absorption at $+$100--130 \kms\ and frequently weaker absorption at $+$170--190 \kms.  This phenomenon varies over the face of the galaxy and is most apparent in sight lines in the southwestern end of the bright ``bar'' on the western side of the galaxy.  The northeastern end of the bar is marked by sight lines with either only one absorption component, or two separated by a smaller velocity.  Sight lines in the ``wing'' (the fainter northwest to southeast extension from the bar on the eastern side of the galaxy) typically have only the lower velocity component.

Clearly, the $+$170--190 \kms\ {\it FUSE} absorption can be identified with the weak $+$182 \kms\ \HI\ feature.  We interpret the $+$100--130 \kms\ absorption component as an unresolved blend of the $+$105 and $+$125 \kms\ absorption features seen in \HI.  Some {\it FUSE} sight lines show distinct absorption at $+$156 \kms\ (good examples are AV378, AV423, and Sk188; see Figures~90, 91, and 96) but it is likely that many other sight lines have weak absorption at this velocity blended with the stronger $+$125 \kms\ feature.  

Since the lower velocity feature is strong in all SMC sight lines, the $+$125 \kms\ component must lie in front of the bulk of the SMC.  The stars must either be immersed at great depth in the lower velocity gas, or located entirely behind it.  The $+$182 \kms\ complex is frequently but not always present and thus may lie behind the lower velocity material.  This interpretation is consistent with the observations of \citet{Cohen84} and \citet{Songalia86}.  Alternatively, this high velocity material could represent patchy, infalling material in front of the SMC \citep{Fitzpatrick85}.

\citet{Songalia86} find \ion{Ca}{2} absorption from the $+$156 \kms\ complex in many of the same sight lines examined here.  While the evidence for emission at this velocity in \HI\ is incontrovertible, we detect little \FeII\ absorption from this component.  Thus, we place the $+$156 \kms\ component to the rear of the other material.  The {\it FUSE} target stars are either located in front of or a short distance within this gas layer.

An additional \FeII\ 1144 \AA\ absorption component is seen in nine sight lines through the bright \HII\ region N66 (NGC\,346-6, NGC\,346-4, NGC\,346-3, NGC\,346-1, AV220, AV229, AV232, AV238, and AV243; Figures~74--80, 82, and 84).  The strong component at $+$125 \kms\ as well as a lower velocity component at approximately $+$75 \kms\ appear only in this region with no sign of the high velocity component.  The $+$75 \kms\ feature appears strongest in sight lines in and near the bright cluster NGC\,346 and becomes noticably weaker in the peripheral sight lines.  Since the $+$75 \kms\ feature is well correlated with N66, we conclude that this gas is associated with the region and is not an intermediate or high velocity cloud.  This low velocity gas may be a $\sim$50 \kms\ outflow of warm gas from ongoing star formation in the region or the remains of an old SNR \citep{Hoopes01a}.

\subsection{LMC Rotation Revealed by Ionic Absorption}
The dynamics of the LMC as a whole can be probed by looking at absorption across the face of the galaxy.  The \PII\ $\lambda$1152 line tends to be the narrowest in our sample, is least likely to be saturated, and best probes the highest column density disk material; with it we can obtain precise velocity measurements of warm gas absorption.  Since the LMC has a large angular extent, a transverse velocity correction was made assuming $\rm v_{tr}=293$ \kms\ at a position angle of $97^\circ$ \citep{Jones94}.  For cases in which multiple absorption components were seen at LMC velocities, an equivalent width-weighted average was calculated.  We find that \PII\ absorption varies considerably both in strength and velocity over the face of the LMC from $\sim$$+$200 to $\sim$$+$300 \kms\ with $<v>=+252$ \kms, $\sigma=23$ \kms.

Figure~3 of \citet{Kim98} shows an \HI\ map with velocity coded by color.  The rotation of the LMC is clockwise with the eastern side closer than the western side.  Thus, material with higher radial velocity is seen to the northern part of the disk and lower velocity to the south.  Figure~99 shows sight lines coded corresponding to \PII\ absorption velocity in the same sense as was used by Kim et~al: dark shades represent lower velocities than lighter shades.  The kinematic center derived by Kim et~al ($\alpha=5^h17.6^m$, $\delta=-69^h02^m$, J2000) is marked by a cross.  The dotted outline is an IRAS 100$\mu$m contour at $\rm 8~MJy~sr^{-1}$.  The \PII\ velocity field is not as well defined as in \HI, but it is clear that the lower velocity material is more prevalent in the south while higher velocity material is found in the north.  The significant scatter in Figure~99 arises in the point-sampled nature of our measurements (versus a larger angular resolution for a radio array).  Furthermore, \HI\ emission line data probes the entire disk whereas the absorption line data looks into the disk to varying depths governed by the positions of the stars themselves.

\section{Summary}
In this work we have presented a set of 94 {\it FUSE} observations of interstellar absorption toward the Magellanic Clouds.  For each sight line, twelve spectral segments are shown highlighting eight different ions and six molecular hydrogen transitions.  \Ha\ images show the distribution of stars and warm ionized gas around each sight line and present a context for interpretation of the {\it FUSE} data.  Some initial findings from this large dataset include the following:

\newcounter{Lcount}
\begin{list}{\arabic{Lcount} --}{\usecounter{Lcount}}

\item \OVI\ absorption is seen along most, if not all, sight lines at Galactic and Magellanic Cloud velocities.  In the case of the LMC, the strength of \OVI\ absorption is not obviously correlated with LMC morphology and it appears at lower velocities than other ions (e.g., \FeII).  See also \citet{Howk01}.  The SMC shows more of a correlation between \OVI\ column density and morphology and the absorption appears closer to the systemic velocity.  See also \citet{Hoopes01b}.

\item Many sight lines toward the LMC show weak absorption at $\sim$$+$50 and/or $\sim$$+$100 \kms\ which we interpret as intermediate and high velocity clouds in the Galactic halo.  These features appear in narrow velocity ranges and are clearly distinct from typical LMC absorption.  The equivalent widths of these features (as measured in \FeII\ $\lambda$1144.938) are not correlated with LMC morphology nor with each other.  In general, the IVC feature appears in eastern sight lines.  The HVC feature is more patchy but also appears concentrated in the east.

\item The SMC frequently shows a bimodal absorption structure possibly arising from expanding superbubbles.  We see multiple components in nearly all sight lines; all sight lines have a strong feature at v$\approx$$+$100--130 \kms\ which we attribute to a blend of absorption from gas associated with from the strong $+$125 \kms\ and weaker $+$105 \kms\ \HI\ complexes with occasional contribution from the $+$156 \kms\ \HI\ component.  Many sight lines--particularly those in the south western portion of the Bar--also show absorption at $+$170--190 \kms\ which arises in the $+$182 \kms\ \HI\ complex.  Sight lines toward the bright \ion{H}{2} region N66 show the strong, low velocity component and an additional component at still lower velocities.

\item The \PII\ $\lambda$1152.818 line is the narrowest ionic line in our sample and traces only the highest column densities of warm disk gas.  We use it to trace the rotation of warm gas in the LMC and find results consistent with existing \HI\ data.

\end{list}

The {\it FUSE} Team deserves recognition for their admirable job acquiring and processing this remendously rich data set.  We are grateful to You-Hua Chu, Michael Wolff and Bart Wakker for contributing {\it FUSE} data to this work; R. Chris Smith for the use of his emission-line filter set; and CTIO for use of their facilities in obtaining the Schmidt images.  The wide-field \Ha\ images the Magellanic Clouds were provided by the Southern H-Alpha Sky Survey Atlas (SHASSA), which is supported by the National Science Foundation.  Charles Hoopes, Scott Friedman, Mark Seibert and Eric Burgh provided essential suggestions and commentary throughout the project.  This research has made use of NASA's Astrophysical Data System and SkyView software and the CDS SIMBAD database.


\begin{figure}
\epsscale{1}\plotone{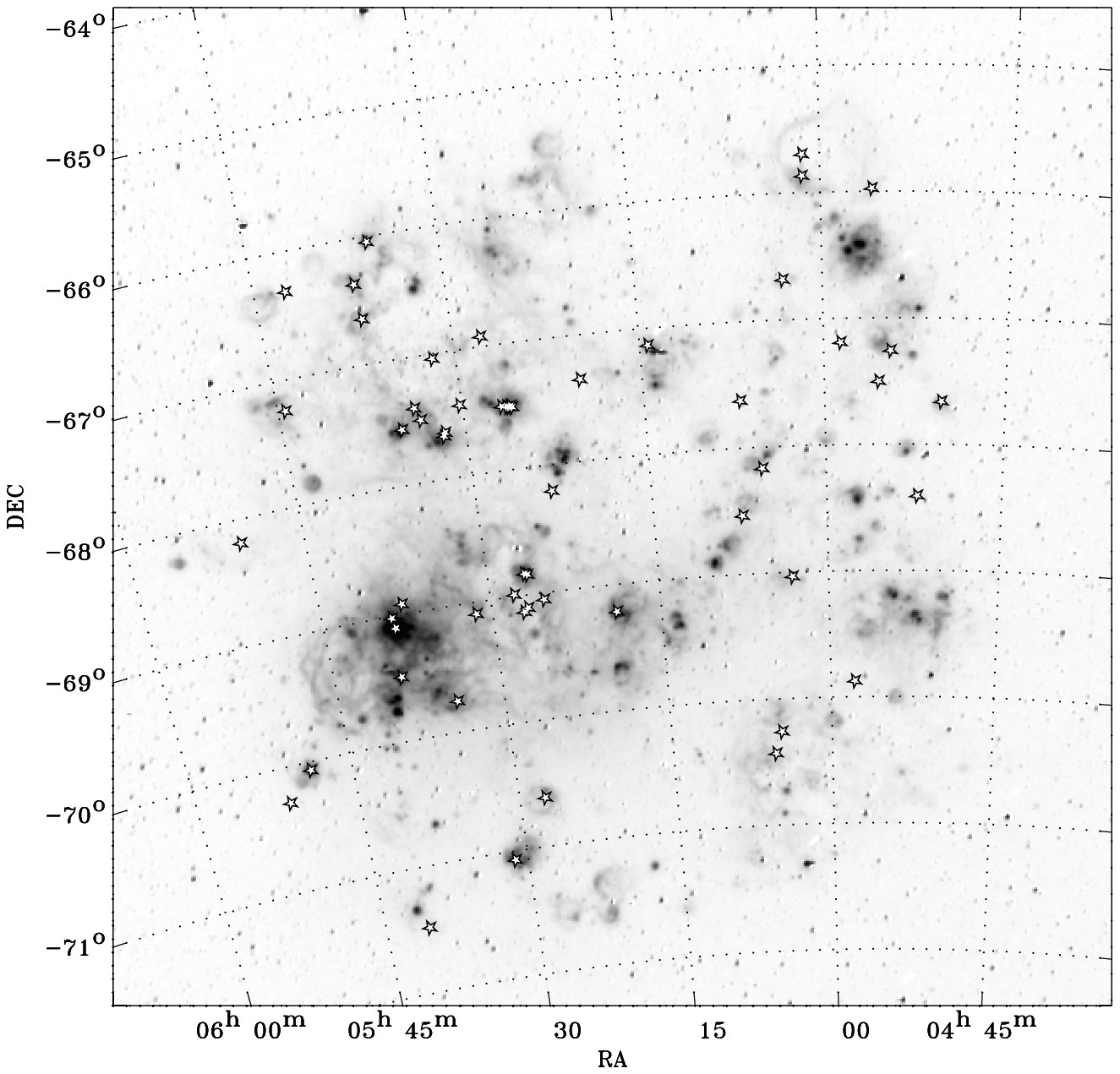}
\caption{LMC pointing map.  Stars represent {\it FUSE} sight lines overlayed on a continuum-subtracted \Ha\ image \citep{SHASSAref}.}
\end{figure}

\begin{figure}
\epsscale{1}\plotone{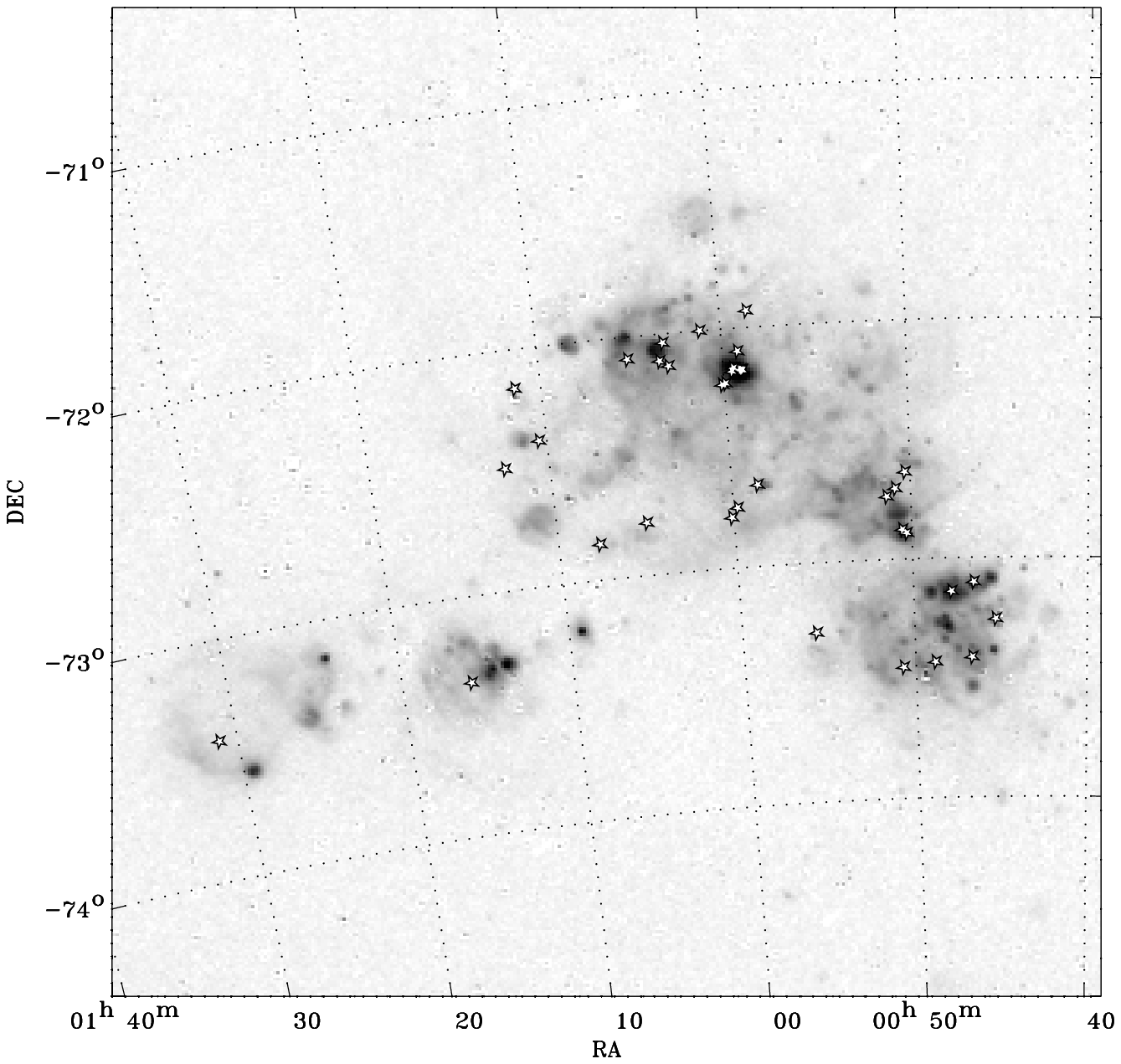}
\caption{SMC pointing map.  Stars represent {\it FUSE} sight lines overlayed on a continuum-subtracted \Ha\ image \citep{SHASSAref}.}
\end{figure}

\clearpage\begin{figure}\figcaption{-96 are available at {\tt http://fuse.pha.jhu.edu/$\sim$danforth/atlas/}}\end{figure}
\setcounter{figure}{96}
\begin{figure}
\epsscale{.75}\plotone{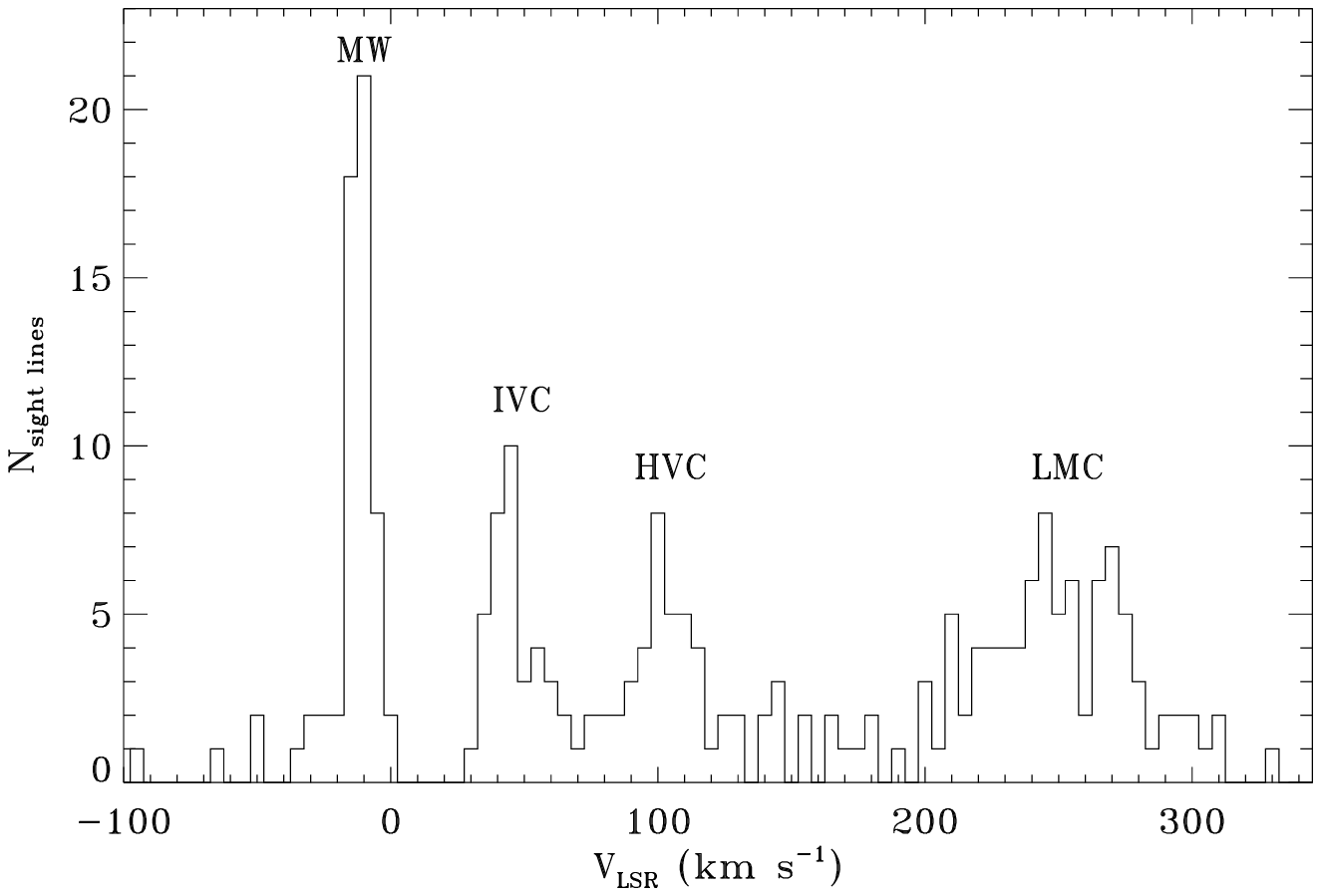}
\figcaption{A histogram of velocity components in \FeII\ $\lambda$1144.938 toward the LMC.  Several distinct populations are apparent.  Milky Way absorption is clearly seen near $-$10 \kms\ while LMC absorption stretches from $+$200 to $+$300 \kms.  A population of absorption features around $+$50 \kms\ shows an intermediate velocity cloud in the Galactic halo.  Similarly, another set of absorption features near $+$100 \kms\ arises in a high velocity cloud between the Galaxy and the LMC.}
\end{figure}

\begin{figure}
\epsscale{1.1}\plottwo{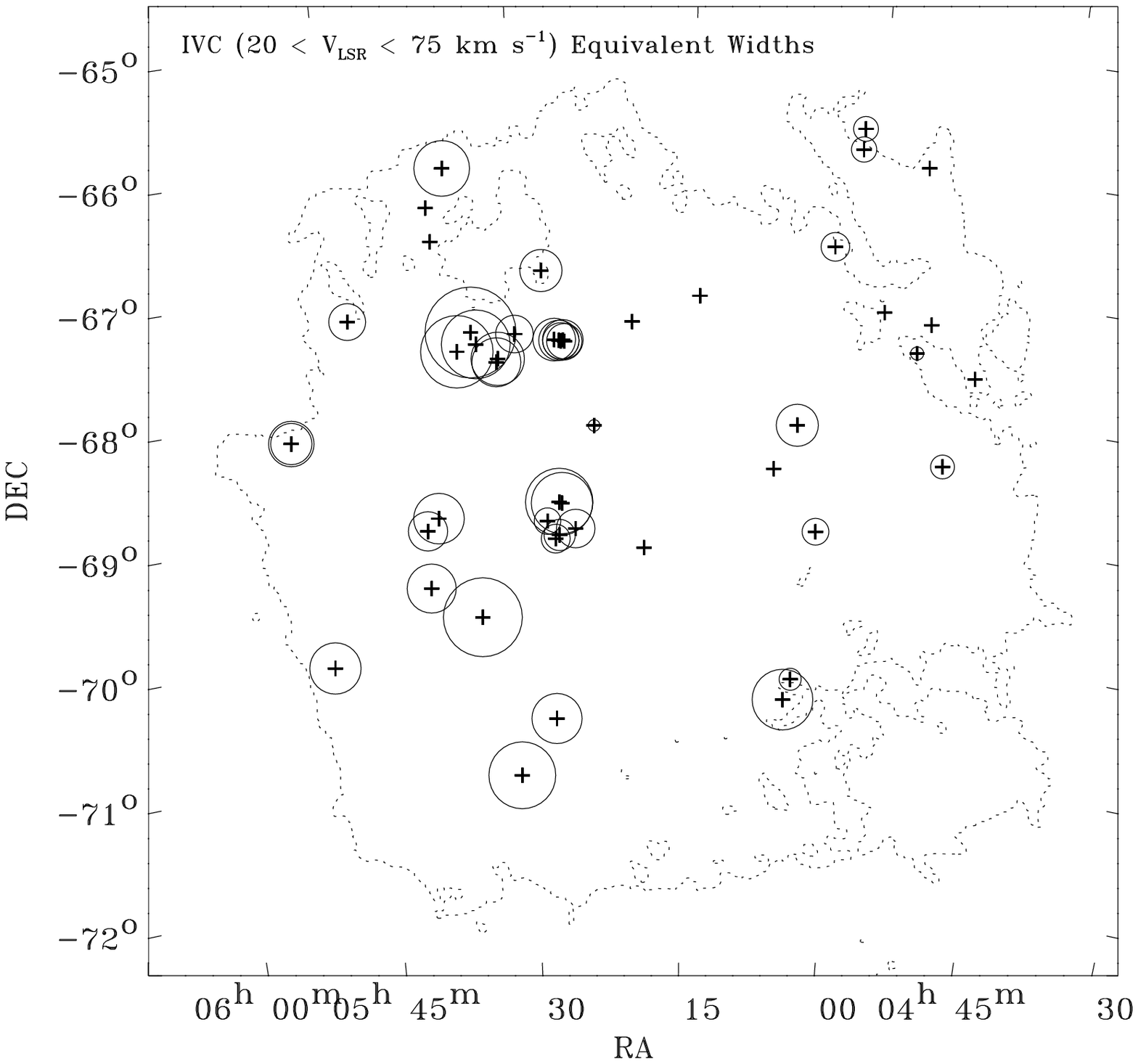}{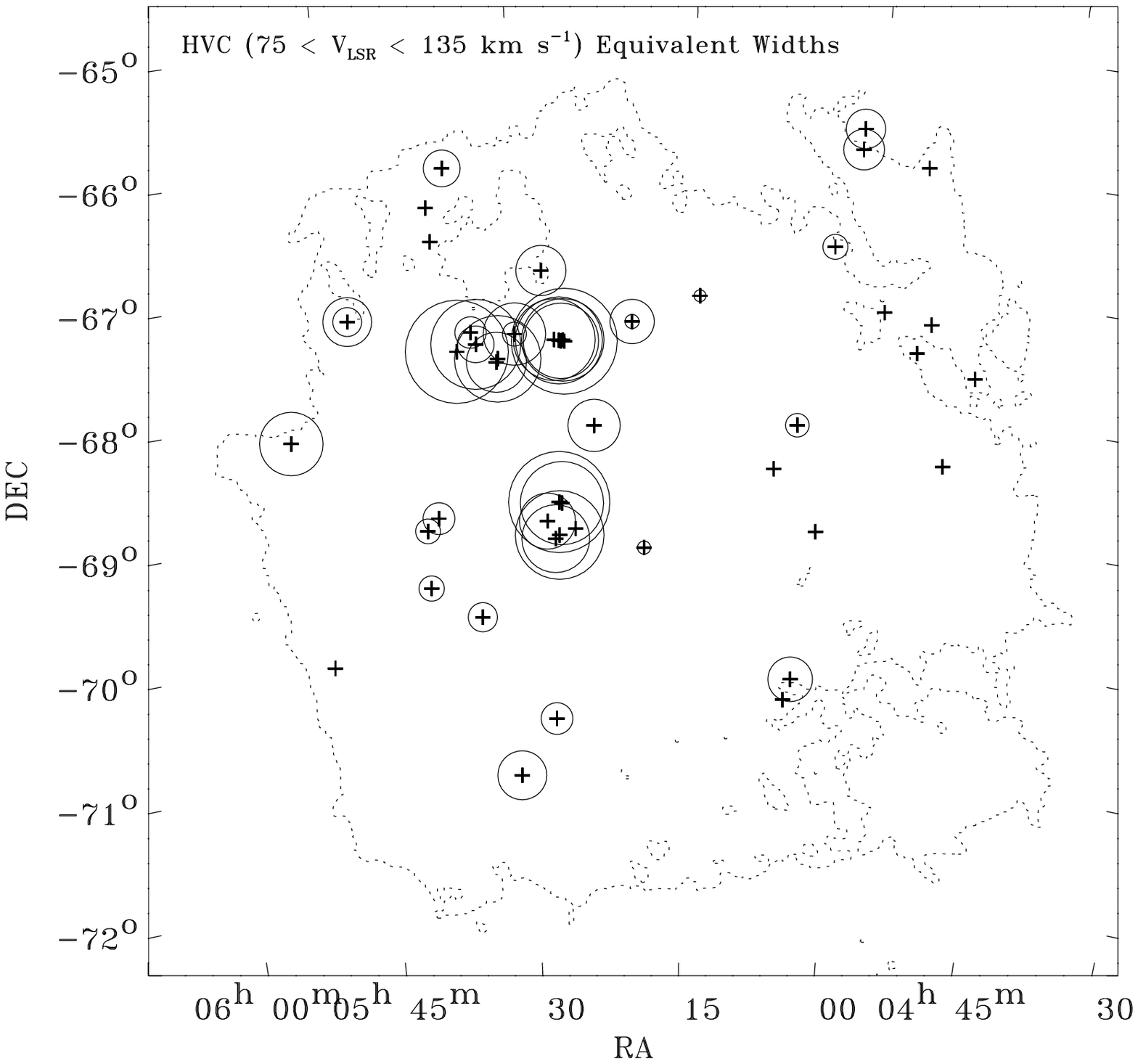}
\figcaption{The distribution of IVC (left panel) and HVC (right panel) absorption features toward the LMC.  Sight lines used in this analysis are marked with crosses.  If an IVC or HVC component was detected, the plus symbol is circled with the radius of the circle being proportional to the measured equivalent width in \FeII$\lambda$1145.  An IRAS 100$\mu$m contour at $\rm 8~MJy~sr^{-1}$ is provided for reference.  Ambiguous and otherwise questionable detections were confirmed qualitatively in \OI$\lambda1039$.  \FeII$\lambda$1145 is a strong transition that shows weak components well and is uncontaminated by other strong lines.  IVC features are clearly found in most sight lines east of about $\rm5^h23^m$ and less commonly to the west.  HVC distribution is more patchy but seems concentrated in the north-east of the galaxy.  Some sight lines have multiple IVC or HVC components and are denoted by multiple circles surrounding a plus symbol.  The smallest circles represent an equivalent width of $\sim$10 m\AA\ and the largest correspond to $\sim$100 m\AA.}
\end{figure}

\begin{figure}
\epsscale{.75}\plotone{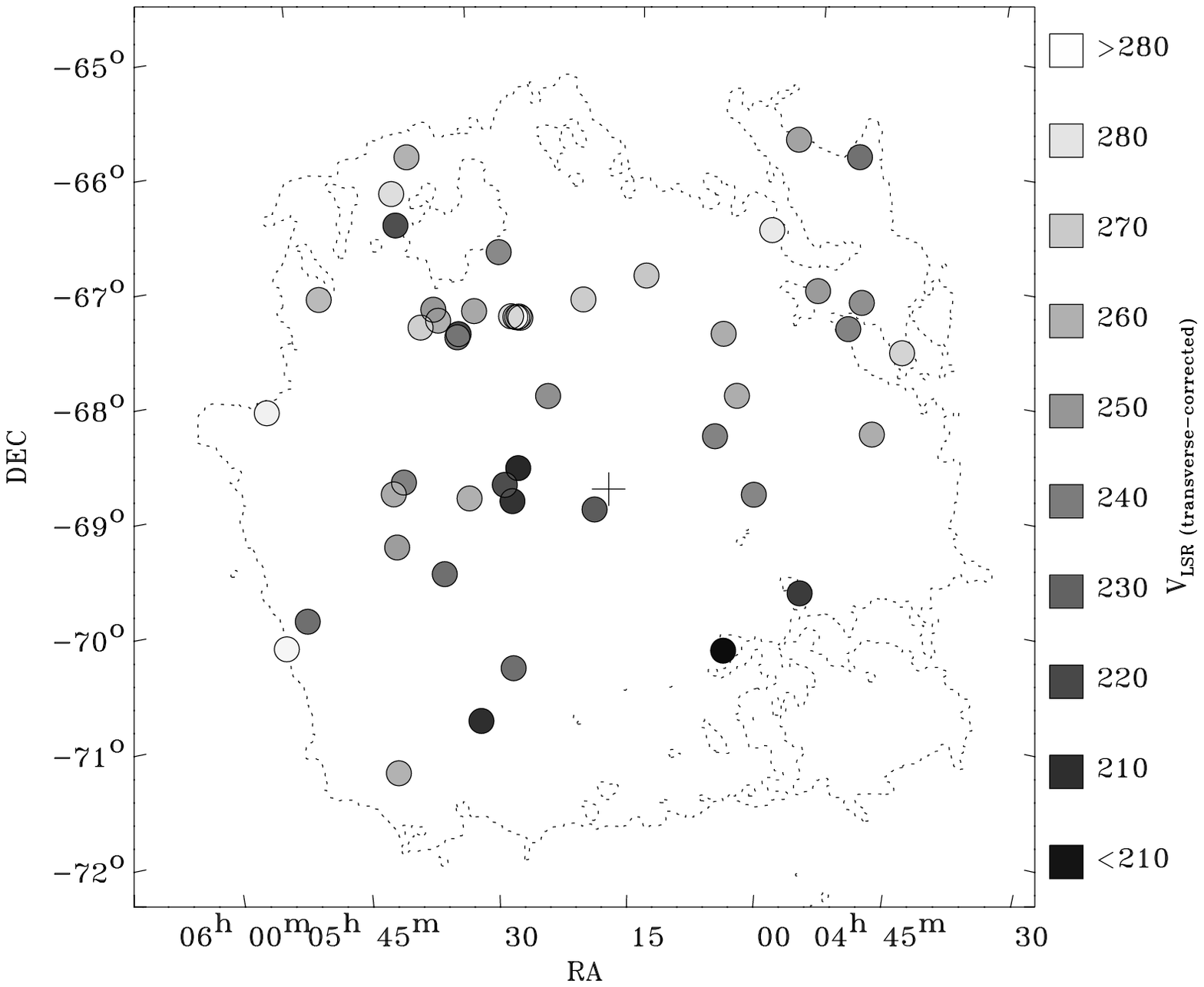}
\figcaption{The velocity field of LMC \PII\ absorption, corrected for the transverse velocity of the LMC, overlayed on an IRAS 100$\mu$m contour at $\rm 8~MJy~sr^{-1}$.  Shades of grey correspond to centroided absorption velocities.  The kinematic center from \citet{Kim98} is marked by a cross.  Lower velocity material (dark) is generally concentrated to the south while higher velocity material (light) is seen more frequently in the north.  Material near the systemic velocity is a medium grey.  This is largely consistent with the \HI\ velocity field of Kim et~al (their Figure~3) though conciderable scatter is present in the \PII\ absorption velocities.}
\end{figure}

\newpage








\begin{deluxetable}{ccll}
\tablecolumns{4}
\tablecaption{FUSE Program Contributions}
\tablehead{\colhead{Program}&\colhead{Targets}&\colhead{PI}&\colhead{Program Title}}
\startdata
P117     & 46 & Hutchings & Hot Stars\\
P103/P203& 22 & Sembach   & Hot Gas in the Galaxy and Magellanic Clouds\\
A049     &  9 & Wakker    & $10^5$K Gas in the Halo of the LMC\\
P115     &  7 & Shull     & Diffuse Molecular Hydrogen\\
P217     &  5 & Fullerton & Less Luminous Hot Stars in the LMC and SMC\\
P221     &  2 & Hutchings & The FUV Extinction Curve in the SMC\\
A111     &  2 & Chu       & Interfaces in N51D\\
A118     &  1 & Wolff     & FUV Nature and Role of Small Dust Grains\\  
A133     &  1 & Fullerton & FUV Diagnostics of Structure in Hot-Star Winds\\
\enddata
\end{deluxetable}

\begin{deluxetable}{lllllllll}
\tabletypesize{\footnotesize}
\tablecolumns{7} 
\tablewidth{0pt} 
\tablecaption{Stellar Parameters for FUSE Targets}
\tablehead{\colhead{Target}                                    &
           \colhead{RA\,(J2000)}                               &
	   \colhead{Dec\,(J2000)}                              & 
           \colhead{Spectral Type}                             &
	   \colhead{Ref.\,\tablenotemark{a}}                   &
           \colhead{$V$}                                       &
           \colhead{(\bv)}                                     &
	   \colhead{Ref.\,\tablenotemark{b}}                   &
           \colhead{Notes}                                    \\
	   \colhead{}                                          &        
	   \colhead{{h}\phn{m}\phn{s}}                         &
	   \colhead{\phn{\arcdeg}~\phn{\arcmin}~\phn{\arcsec}} &
	   \colhead{}                                          & 	       
	   \colhead{}                                          & 	    
	   \colhead{}                                          & 
	   \colhead{}                                          & 
	   \colhead{}                                          & 
	   \colhead{}                                          }
\startdata
\cutinhead{Large Magellanic Cloud }
\sk{65}{21}  & 05 01 22.33 & $-$65 41 48.1 & O9.7 Iab      & W95  &    12.02     & $-$0.16 & I75 &                   \\
\sk{65}{22}  & 05 01 24.00 & $-$65 52 00.0 & O6 Iaf+       & W77  &    12.07     & $-$0.19 & I79 & HDE\,270952       \\
\sk{66}{18}  & 04 55 59.88 & $-$65 58 30.0 & O6 V((f))     & M95  &    13.50     & $-$0.20 & I79 &                   \\
\sk{66}{51}  & 05 03 10.20 & $-$66 40 54.0 & WN8h          & S96  &    12.71     & $-$0.23 & F83 & Br\,13, HD\,33133 \\
\sk{66}{100} & 05 27 45.59 & $-$66 55 15.0 & O6 II(f)      & W95  &    13.26     & $-$0.21 & I79 &                   \\
BI\,229      & 05 35 32.20 & $-$66 02 37.6 & O7 V-III      & W01b &    12.95     & $-$0.17 & B75 &                   \\
\sk{66}{169} & 05 36 54.50 & $-$66 38 25.0 & O9 .7Ia+      & F88  &    11.56     & $-$0.13 & R78 & HDE\,269889\\
\sk{66}{172} & 05 37 05.56 & $-$66 21 35.7 & O2 III(f$^\ast$)+OB&W01a& 13.13     & $-$0.12 & R78 &                   \\
\sk{66}{185} & 05 42 30.00 & $-$66 19 00.0 & B0 Iab        & C86  &    13.11     & $-$0.19 & I79 &                   \\
\sk{67}{05}  & 04 50 18.96 & $-$67 39 37.9 & O9.7 Ib       & W97  &    11.34     & $-$0.12 & A72 & HDE\,268605       \\
\sk{67}{14}  & 04 54 31.92 & $-$67 15 24.9 & B1.5 Ia       & F88  &    11.52     & $-$0.10 & A72 & HDE\,268685       \\
\sk{67}{20}  & 04 55 31.50 & $-$67 30 01.0 & WN4b          & S96  &    13.86     & $-$0.28 & F83 & HD\,32109         \\
\sk{67}{28}  & 04 58 36.84 & $-$67 11 22.7 & B0.7 Ia       & F88  &    12.28     & $-$0.14 & I82 &                   \\
\sk{67}{46}  & 05 07 01.62 & $-$67 37 29.6 & B1.5 I        & J01  &    12.34     & $-$0.06 & I75 &                   \\
\sk{67}{69}  & 05 14 20.16 & $-$67 08 03.5 & O4 III(f)     & G87a &    13.09     & $-$0.16 & I79 &                   \\
\sk{67}{76}  & 05 20 05.81 & $-$67 21 08.9 & B0            & R78  &    12.42     & $-$0.13 & I75 &                   \\
\sk{67}{101} & 05 25 56.36 & $-$67 30 28.7 & O8 II((f))    & W01b &    12.63     & $-$0.17 & I75 & BI\,169           \\
\sk{67}{104} & 05 26 04.00 & $-$67 29 48.0 & WC4(+O?)+O8I: & M90  &    11.44     & $-$0.17 & F83 & Br\,31, HD\,36402 \\
\sk{67}{106} & 05 26 13.76 & $-$67 30 15.1 & B0:           & R78  &    11.78     & $-$0.17 & R78 & HDE\,269525\\
\sk{67}{107} & 05 26 24.00 & $-$67 30 00.0 & B0            & R78  &    12.50     & $-$0.12 & I75 &                   \\
\sk{67}{111} & 05 26 48.00 & $-$67 29 33.0 & O6: Iafpe     & W01b &    12.57     & $-$0.20 & I75 &                   \\
\sk{67}{144} & 05 30 12.22 & $-$67 26 08.4 & WC4           & T88  &    13.60     & $-$0.31 & F83 & HD\,37026\\
\sk{67}{166} & 05 31 44.31 & $-$67 38 00.6 & O4 If+        & W77  &    12.27     & $-$0.22 & A72 & HDE\,269698 \\
\sk{67}{167} & 05 31 51.98 & $-$67 39 41.1 & O4 Inf+       & G87a &    12.54     & $-$0.19 & I75 &                   \\
\sk{67}{169} & 05 31 51.68 & $-$67 02 22.3 & B1 Ia         & F88  &    12.18     & $-$0.12 & I75 &                   \\
\sk{67}{191} & 05 33 34.12 & $-$67 30 19.6 & O8 V          & C86  &    13.46     & $-$0.21 & I79 &                   \\
BI\,208      & 05 33 57.45 & $-$67 24 20.0 & O7 Vz         & W01b &    13.96     & $-$0.24 & I82 &                   \\
\sk{67}{211} & 05 35 13.92 & $-$67 33 27.0 & O2 III(f$^\ast$)&W01a&    12.28     & $-$0.23 & A72 & HDE\,269810       \\
BI\,272      & 05 44 23.18 & $-$67 14 29.3 & O7: III-II:   & W01b &    13.28     & $-$0.22 & I82 &                   \\
\sk{68}{03}  & 04 52 15.56 & $-$68 24 26.9 & O9 I          & C86  &    13.13     & $-$0.13 & I79 &                   \\
\sk{68}{41}  & 05 05 27.20 & $-$68 10 02.7 & B0.5 Ia       & F88  &    12.01     & $-$0.14 & I82 &                   \\
\sk{68}{52}  & 05 07 20.60 & $-$68 32 09.6 & B0 Ia         & W77  &    11.54     & $-$0.07 & A72 & HDE\,269050       \\
\sk{68}{75}  & 05 23 28.52 & $-$68 12 22.8 & B1 I          & J01  &    12.03     & $-$0.06 & A72 & HDE\,269463       \\
\sk{68}{80}  & 05 26 30.43 & $-$68 50 26.6 & WC4+O6V-III   & M90  &    12.42     & $-$0.23 & F83 & HD\,36521         \\
\sk{68}{82}  & 05 26 45.30 & $-$68 49 52.8 & WN5?b+(B3I)   & S96  & \phn9.86     & $-$0.03 & I82 & Br\,34, HDE\,269546\\
\sk{68}{135} & 05 37 48.60 & $-$68 55 08.0 & ON9.7 Ia+     & W77  &    11.36     &\phs0.00 & A72 & HDE\,269896       \\
\sk{68}{171} & 05 50 22.86 & $-$68 11 18.4 & B0.7 Ia       & F91  &    12.01     & $-$0.09 & A72 & HDE\,270220       \\
\sk{69}{52}  & 04 57 48.50 & $-$69 52 22.0 & B2 Ia         & F88  &    11.50     & $-$0.03 & A72 & HDE\,268867       \\
\sk{69}{59}  & 05 03 12.00 & $-$69 02 00.0 & B0            & R78  &    12.13     & $-$0.12 & I75 & HDE\,268960       \\
\sk{69}{104} & 05 18 59.57 & $-$69 12 54.7 & O6 Ib(f)      & Wpc  &    12.10     & $-$0.21 & A72 & HDE\,269357       \\
\sk{69}{124} & 05 25 18.37 & $-$69 03 11.1 & O9 Ib         & C86  &    12.81     & $-$0.18 & I82 &                   \\
BI\,170      & 05 26 47.79 & $-$69 06 11.7 & O9.5 Ib       & W01b &    13.09     & $-$0.17 & B75 &                   \\
BI\,173      & 05 27 10.08 & $-$69 07 56.2 & O8 II:        & W01b &    13.00     & $-$0.14 & B75 &                   \\
\sk{69}{142}a& 05 27 52.75 & $-$68 59 08.6 & WN10h         & C97  &    11.88\tnc & $-$0.04 & S86 & BE294, HDE\,269582\\
\sk{69}{175} & 05 31 25.61 & $-$69 05 38.4 & WN11h         & C97  &    11.90     & $-$0.07 & I75 & S119, HDE\,269687 \\
\sk{69}{191} & 05 34 19.39 & $-$69 45 10.0 & WC4           & T88  &    13.35     & $-$0.20 & F83 & HD\,37680         \\
Mk\,42       & 05 38 42.10 & $-$69 05 54.7 & O3 If$^\ast$/WN6-A&W97&   10.96\tnd & $+$0.12 & F83 & Br\,77            \\
\sk{69}{243} & 05 38 42.57 & $-$69 06 03.2 & WN5+OB        & M98  & \phn9.50\tnd & $+$0.13 & F83 & R136, HD\,38268\\
\sk{69}{246} & 05 38 54.00 & $-$69 01 00.0 & WN6h          & S96  &    11.16     & $-$0.16 & F83 & Br89,HD\,38282 \\
\sk{69}{249}c& 05 38 58.25 & $-$69 29 19.1 & WN9h          & S96  &    12.63     & $-$0.13 & T98 & HDE\,269927c      \\
\sk{70}{60}  & 05 04 40.94 & $-$70 15 34.5 & O4-5 V:n      & Wpc  &    13.85     & $-$0.19 & R78 &                   \\
\sk{70}{69}  & 05 05 18.73 & $-$70 25 49.8 & O5 V          & W95  &    13.94     & $-$0.27:& R78 &                   \\
\sk{70}{91}  & 05 27 33.74 & $-$70 36 48.3 & O6.5 V        & C86  &    12.78     & $-$0.23 & I79 &                   \\
\sk{70}{115} & 05 48 49.76 & $-$70 03 57.5 & O6.5 Iaf      & Wpc  &    12.24     & $-$0.10 & I75 & HDE\,270145       \\
\sk{70}{120} & 05 51 20.85 & $-$70 17 08.7 & B1 Ia         & F88  &    11.59     & $-$0.06 & A72 & HDE\,270196       \\
\sk{71}{45}  & 05 31 15.55 & $-$71 04 08.9 & O4-5 III(f)   & W77  &    11.51\tnd & $-$0.19 & H91 & HDE\,269676       \\
\sk{71}{50}  & 05 40 43.32 & $-$71 28 59.3 & O6.5 III      & C86  &    13.44     & $-$0.12 & R78 &                   \\
\cutinhead{Small Magellanic Cloud}
AV\,6        & 00 45 18.20 & $-$73 15 23.4 & O9 III        & L97  &    13.46     & $+$0.03 & A75 &                   \\
AV\,14       & 00 46 32.66 & $-$73 06 05.6 & O3-4 V        & G87b &    13.77     & $-$0.19 & A75 & Sk\,9             \\
AV\,15       & 00 46 42.19 & $-$73 24 54.7 & O6.5 II(f)    & W00  &    13.17     & $-$0.21 & I78 & Sk\,10            \\
AV\,26       & 00 47 50.07 & $-$73 08 20.7 & O7 III        & G87b &    12.55     & $-$0.20 & A75 & Sk\,18            \\
AV\,47       & 00 48 51.35 & $-$73 25 57.6 & O8 III((f))   & W00  &    13.38     & $-$0.26 & A75 &                   \\
AV\,69       & 00 50 17.40 & $-$72 53 29.9 & OC7.5 III((f))& W00  &    13.35     & $-$0.22 & A75 &                   \\
AV\,70       & 00 50 18.14 & $-$72 38 09.8 & O9.5 Iw       & W83  &    12.38     & $-$0.17 & A75 & Sk\,35            \\
AV\,75       & 00 50 32.50 & $-$72 52 36.2 & O5 III(f+)    & W00  &    12.79     & $-$0.16 & I78 & Sk\,38            \\
AV\,81       & 00 50 43.47 & $-$73 27 06.1 & WN5h          & S96  &    13.29     & $-$0.10 & A75 & Sk\,41, AB4       \\
AV\,83       & 00 50 52.01 & $-$72 42 14.5 & O7 Iaf+       & W00  &    13.58     & $-$0.13 & W00 &                   \\
AV\,95       & 00 51 21.54 & $-$72 44 12.9 & O7 III((f))   & W00  &    13.91     & $-$0.30 & A75 &                   \\
AV\,170      & 00 55 42.48 & $-$73 17 30.0 & O9.7 III      & W00  &    14.09     & $-$0.23 & A75 &                   \\
AV\,207      & 00 58 33.19 & $-$71 55 46.5 & O7 V          & C82  &    14.37     & $-$0.22 & A75 &                   \\
AV\,208      & 00 58 33.18 & $-$72 39 31.6 & O8 V          & M95  &    14.10     & $+$0.01 & A75 &                   \\
NGC\,346-6   & 00 58 57.74 & $-$72 10 33.6 & O4 V((f))     & W95  &    14.02     & $-$0.24 & M89 & MPG-324           \\
NGC\,346-4   & 00 59 00.39 & $-$72 10 37.9 & O5-6 V        & W86  &    13.66     & $-$0.23 & M89 & MPG-342           \\
NGC\,346-3   & 00 59 01.09 & $-$72 10 28.2 & O2 III(f$^\ast$)&W01a&    13.50     & $-$0.23 & M89 & MPG-355           \\
NGC\,346-1   & 00 59 04.81 & $-$72 10 24.8 & O4 III(n)(f)  & W86  &    12.57     & $-$0.20 & M89 & MPG-435           \\
AV\,220      & 00 59 10.11 & $-$72 05 48.1 & O6.5f?p       & W00  &    14.50     & $-$0.22 & A75 &                   \\
AV\,229      & 00 59 26.55 & $-$72 09 53.8 & WN var        & K94  &    11.86\tnc & $-$0.26 & A75 & HD\,5980, Sk\,78  \\
AV\,232      & 00 59 31.95 & $-$72 10 45.8 & O7 Iaf+       & W77  &    12.36     & $-$0.20 & A75 & Sk\,80            \\
AV\,235      & 00 59 45.72 & $-$72 44 56.1 & B0 Iaw        & W83  &    12.20     & $-$0.18 & A75 & Sk\,82            \\
AV\,238      & 00 59 55.61 & $-$72 13 37.7 & O9.5 III      & W00  &    13.77     & $-$0.22 & A75 &                   \\
AV\,242      & 01 00 06.84 & $-$72 13 57.0 & B0.7 Iaw      & W83  &    12.11     & $-$0.13 & A75 & Sk\,85            \\
AV\,243      & 01 00 06.80 & $-$72 47 19.0 & O6 V          & W95  &    13.87     & $-$0.22 & A75 & Sk\,84            \\
AV\,264      & 01 01 07.72 & $-$71 59 58.6 & B1 Ia         & L97  &    12.36     & $-$0.15 & A75 & Sk\,94            \\
AV\,321      & 01 02 57.04 & $-$72 08 09.3 & B0 IIIww      & G87b &    13.88     & $-$0.21 & A75 &                   \\
AV\,327      & 01 03 10.58 & $-$72 02 13.8 & O9.5 II-Ibw   & W00  &    13.25     & $-$0.22 & A79 &                   \\
AV\,332      & 01 03 25.24 & $-$72 06 43.3 & O6.5(n)+WN3   & W77  &    12.41     & $-$0.29 & A75 & Sk\,108           \\
AV\,372      & 01 04 55.73 & $-$72 46 47.7 & O9.5 Iabw     & W01b &    12.63     & $-$0.18 & A75 & Sk\,116           \\
AV\,378      & 01 05 09.44 & $-$72 05 35.0 & O8 V          & G87b &    13.88     & $-$0.24 & A75 &                   \\
AV\,423      & 01 07 40.43 & $-$72 50 59.6 & O9.5 II(n)    & W01b &    13.28     & $-$0.19 & A75 & Sk\,132           \\
AV\,451      & 01 10 25.96 & $-$72 23 28.3 & O9 V          & G87b &    14.15     & $-$0.23 & A75 &                   \\ 
AV\,461      & 01 11 25.49 & $-$72 09 48.5 & O8 V          & G87b &    14.66     & $-$0.31 & I78 &                   \\
AV\,469      & 01 12 28.96 & $-$72 29 28.8 & O8.5 II((f))  & W01b &    13.20     & $-$0.22 & A75 & Sk\,148           \\
AV\,488      & 01 15 58.84 & $-$73 21 24.1 & B0.5 Iaw      & W83  &    11.89     & $-$0.13 & A77 & Sk\,159           \\
Sk\,188      & 01 31 04.23 & $-$73 25 02.2 & WO3 + O4V     & C98  &    12.41     & $-$0.29 & A75 &                   \\
\enddata

\tablenotetext{a}{References for Spectral Types.~
  C82  = \citealt{Crampton82};
  C86  = \citealt{Conti86};
  C97  = \citealt{Crowther97};
  C98  = \citealt{Crowther98};
  F88  = \citealt{Fitzpatrick88};
  F91  = \citealt{Fitzpatrick91};  
  G87a = \citealt{Garmany87a};
  G87b = \citealt{Garmany87b};
  J01  = \citealt{Jaxon01};
  K94  = \citealt{Koenigsberger94}; 
  L97  = \citealt{Lennon97};
  M90  = \citealt{Moffat90};
  M95  = \citealt{Massey95};
  M98  = \citealt{Massey98};
  R78  = \citealt{Rousseau78};
  S96  = \citealt{Smith96};
  T88  = \citealt{Torres88};
  W77  = \citealt{Walborn77};
  W82  = \citealt{Walborn82};
  W83  = \citealt{Walborn83};  
  W86  = \citealt{Walborn86};
  W95  = \citealt{Walborn95};
  W97  = \citealt{Walborn97};
  W00  = \citealt{Walborn00};
  W01a = \citealt{Walborn01a};
  W01b = \citealt{Walborn01b};
  Wpc  = Walborn, private communication.           }
  
\tablenotetext{b}{References for Photometry.~
  A72  = \citealt{Ardeberg72};
  A75  = \citealt{Azzopardi75};
  A77  = \citealt{Ardeberg77};
  A79  = \citealt{Azzopardi79};
  B75  = \citealt{Brunet75};
  F83  = \citealt{Feitzinger83};
  H91  = \citealt{Heydari91};
  I75  = \citealt{Isserstedt75};
  I78  = \citealt{Isserstedt78};
  I79  = \citealt{Isserstedt79};
  I82  = \citealt{Isserstedt82};
  M89  = \citealt{Massey89};                          
  R78  = \citealt{Rousseau78};
  S86  = \citealt{Stahl86};
  T98  = \citealt{Testor98};
  W00  = \citealt{Walborn00}.
                        }
\tablenotetext{c}{Variable.}

\tablenotetext{d}{Photometric measurements refer to a blend of several nearby sources.}

\end{deluxetable}

\begin{deluxetable}{llcrrll}
\tabletypesize{\footnotesize}
\tablecolumns{7} 
\tablewidth{0pt} 
\tablecaption{FUSE Observational Parameters}
\tablehead{\colhead{Target}	&
	   \colhead{FUSEID}	&
	   \colhead{Obs}	&
	   \colhead{\#Exp}	&
	   \colhead{Exptime}	&
	   \colhead{Obs Date}	&
	   \colhead{Notes}	}
\startdata
\cutinhead{Large Magellanic Cloud }
\sk{65}{21}  &P10309&1,2,3,4& 4& 18126& 2000-10-05&                   \\
\sk{65}{22}  &P10310&2      &11& 27199& 1999-12-20& HDE\,270952       \\
\sk{66}{18}  &A04901&2      & 1&  3228& 2000-09-27&                   \\
\sk{66}{51}  &P11745&1      & 4&  4617& 2000-09-30& Br\,13, HD\,33133 \\
\sk{66}{100} &P11723&3      & 2&  7148& 1999-12-20&                   \\
BI\,229      &P11728&1      & 2&  5623& 2000-02-11&                   \\
\sk{66}{169} &P11738&1      & 3&  5176& 2000-02-11& HDE\,269889       \\
\sk{66}{172} &P11722&1      & 1&  3562& 2000-09-26&                   \\
\sk{66}{185} &A04909&2      & 4&  7022& 2000-02-11&                   \\
\sk{67}{05}  &P10307&3,4    & 2&  7410& 2000-10-07& HDE\,268605       \\
\sk{67}{14}  &P11742&1,2,3  & 4& 14450& 2000-09-27& HDE\,268685       \\
\sk{67}{20}  &P11744&1,2    & 5& 16685& 2000-10-12& HD\,32109         \\
\sk{67}{28}  &A04902&1,2    & 2&  6222& 1999-12-20&                   \\
\sk{67}{46}  &A04915&1      & 4&  3470& 2000-09-26&                   \\
\sk{67}{69}  &P11717&3      & 4&  7793& 1999-12-20&                   \\
\sk{67}{76}  &P10312&1      & 5& 29596& 1999-12-13&                   \\
\sk{67}{101} &P11734&1,3    & 5& 10753& 2000-09-29& BI\,169           \\
\sk{67}{104} &P10313&2      & 3&  5108& 1999-12-17& Br\,31, HD\,36402 \\
\sk{67}{106} &A11101&1      & 2& 11270& 2000-02-09& HDE\,269525       \\
\sk{67}{107} &A11102&2      & 2& 11184& 2000-02-09&                   \\
\sk{67}{111} &P11730&1      & 2&  7998& 2000-02-11&                   \\
\sk{67}{144} &P11750&1      & 3&  8125& 2000-02-12& HD\,37026         \\
\sk{67}{166} &A13301&many   &63&222358& 2000-09-26& HDE\,269698       \\
\sk{67}{167} &P11719&1,2    & 3& 10165& 2000-09-27&                   \\
\sk{67}{169} &P10316&3      & 4& 40544& 1999-12-15&                   \\
\sk{67}{191} &P11731&1,2    & 3& 15029& 2000-09-29&                   \\
BI\,208      &P11727&2,3,4,5& 4& 15635& 2000-10-01&                   \\
\sk{67}{211} &P11716&3      & 7&  8756& 1999-12-20& HDE\,269810       \\
BI\,272      &P11729&1,2    & 1&  6599& 1999-12-18&                   \\
\sk{68}{03}  &A04904&1,2    & 4&  9306& 2000-10-05&                   \\
\sk{68}{41}  &P11741&1,2 & 2& 10885& 2000-10-03&                   \\
\sk{68}{52}  &P11740&1      & 5&  8080& 2000-10-01& HDE\,269050       \\
\sk{68}{75}  &A04905&1,2    & 7&  6494& 2000-10-03& HDE\,269463       \\
\sk{68}{80}  &P10314&2      & 4&  9703& 1999-12-17& HD\,36521         \\
\sk{68}{82}  &P20301&1,2,3,4& 5&  9617& 2000-10-13& Br\,34, HDE\,269546\\
\sk{68}{135} &P11739&1      & 3&  7073& 2000-02-12& HDE\,269896       \\
\sk{68}{171} &A04908&1      & 3&  4577& 2000-02-11& HDE\,270220       \\
\sk{69}{52}  &P11743&1      & 1&  5486& 2000-10-02& HDE\,268867       \\
\sk{69}{59}  &P10311&3      & 4& 25664& 1999-12-15& HDE\,268960       \\
\sk{69}{104} &P11724&1      & 1&  4130& 2000-09-30& HDE\,269357       \\
\sk{69}{124} &P11736&1,2    & 3& 12672& 2000-10-03&                   \\
BI\,170      &P11737&1      & 1&  4280& 2000-09-30&                   \\
BI\,173      &P11732&1,2    & 4& 11353& 2000-10-03&                   \\
\sk{69}{142}a&P11747&1      & 1&  4612& 2000-02-10& BE294, HDE\,269582\\
\sk{69}{175} &P11748&1      & 1&  3780& 2000-10-02& S119, HDE\,269687 \\
\sk{69}{191} &P11751&1      & 3&  7041& 2000-02-12& HD\,37680         \\
Mk\,42       &P11718&2,3,4  & 5& 15337& 2000-09-28& Br\,77            \\
\sk{69}{243} &P10317&3,4,5,6& 4& 17348& 2000-10-04& R136, HD\,38268\\
\sk{69}{246} &P10318&2      & 4& 22077& 1999-12-16& Br89,HD\,38282 \\
\sk{69}{249}c&P11746&1      & 2&  7202& 2000-02-09& HDE\,269927c      \\
\sk{70}{60}  &P11720&1      & 2&  7900& 2000-12-04&                   \\
\sk{70}{69}  &P11721&1      & 2&  6107& 2000-12-04&                   \\
\sk{70}{91}  &P11725&1      & 1&  5491& 2000-10-05&                   \\
\sk{70}{115} &P11726&1      & 2&  5206& 2000-02-12& HDE\,270145       \\
\sk{70}{120} &A04910&1      & 2&  7103& 2000-09-28& HDE\,270196       \\
\sk{71}{45}  &P10315&1,2,3,4& 4& 18859& 2000-10-02& HDE\,269676       \\
\sk{71}{50}  &A04912&1      & 7&  5697& 2000-07-05&                   \\
\cutinhead{Small Magellanic Cloud}
AV\,6        &P22101&1      & 2& 7130& 2001-06-14&                   \\
AV\,14       &P11753&1      & 2& 6770& 2000-07-01& Sk\,9             \\
AV\,15       &P11501&1      &12&14643& 2000-05-30& Sk\,10            \\
AV\,26       &P11760&1      & 1& 4030& 2000-07-02& Sk\,18            \\
AV\,47       &P11502&2      & 5&16258& 2000-05-30&                   \\
AV\,69       &P11503&3      &13&17586& 2000-05-31&                   \\
AV\,70       &A11802&2,3    & 3& 4794& 2000-10-03& Sk\,35            \\
AV\,75       &P11504&4      & 4&14375& 2000-05-31& Sk\,38            \\
AV\,81       &P21708&1      & 2& 9192& 2001-06-14& Sk\,41, AB4    \\
AV\,83       &P11762&1      & 1& 4066& 2000-07-02&                   \\
AV\,95       &P11505&5      &11&14298& 2000-05-31&                   \\
AV\,170      &P21707&1      & 3& 9819& 2001-06-12&               \\
AV\,207      &P11759&1      & 3& 4060& 2000-10-03&                   \\
AV\,208      &P22103&1      & 2& 8431& 2001-06-14&               \\
NGC\,346-6   &P20305&2      & 5&10992& 2001-09-25& MPG-324           \\
NGC\,346-4   &P20304&1      & 6&11853& 2001-08-25& MPG-342           \\
NGC\,346-3   &P20303&1      & 3& 8482& 2001-08-25& MPG-355           \\
NGC\,346-1   &P20302&1      & 3& 4625& 2001-08-25& MPG-435           \\
AV\,220      &P21704&1      & 3&11849& 2001-06-15&               \\
AV\,229      &P10301&1      & 4& 5734& 2000-07-02& HD\,5980, Sk\,78  \\
AV\,232      &P10302&1      & 4&11699& 2000-07-02& Sk\,80            \\
AV\,235      &P10303&1      &10&17203& 2000-07-02& Sk\,82            \\
AV\,238      &P11766&1      & 2&11103& 2000-10-12&                   \\
AV\,242      &P11769&1      & 2& 5063& 2000-10-04& Sk\,85            \\
AV\,243      &P11758&2      & 3& 5255& 2001-06-15& Sk\,84        \\
AV\,264      &P11770&1      & 3& 4494& 2000-10-06& Sk\,94            \\
AV\,321      &P11506&6      & 6&16936& 2000-06-01&                   \\
AV\,327      &P11764&1      & 4& 4690& 2000-10-02&                   \\
AV\,332      &P10304&1      &16&13958& 2000-07-03& Sk\,108           \\
AV\,372      &P11765&1      & 1& 4350& 2000-10-11& Sk\,116           \\
AV\,378      &P11507&7      & 8&14685& 2000-06-01&                   \\
AV\,423      &P11767&1      & 1& 4000& 2000-10-10& Sk\,132           \\
AV\,451      &P21706&1      & 2& 6835& 2001-06-13&               \\ 
AV\,461      &P21705&1      & 3& 7752& 2001-06-13&               \\
AV\,469      &P11763&1      & 3& 8169& 2000-10-11& Sk\,148           \\
AV\,488      &P11768\tablenotemark{a}&1,2,3  & 7&15294& 2000-10-10& Sk\,159           \\
Sk\,188      &P10306&1,2,3,4& 8&23113& 2000-10-10&                   \\
\enddata
\tablenotetext{a}{includes two exposures from P1030501}
\end{deluxetable}

\begin{deluxetable}{lrcl}
\tabletypesize{\footnotesize}
\tablecolumns{4} 
\tablewidth{0pt} 
\tablecaption{Transitions Included in the Atlas}
\tablehead{\colhead{Ion}&\colhead{Wavelength}&\colhead{Ionization Range (eV)\tablenotemark{a}}&\colhead{$f_\lambda$\tablenotemark{c}}}
\startdata
\OVI          & 1031.926 &113.90 -- 138.12& 0.1329\\
\CIII         &  977.020 & 24.383 -- 47.887& 0.7620\\
\CII          & 1036.337 & 11.260 -- 24.383& 0.1231\\
\SiII         & 1020.699 & ~8.151 -- 16.345& 0.02828\\
\PII          & 1152.818 & 10.486 -- 19.725& 0.2361\\
\OI           & 1039.230 & ~0~~~ -- 13.618& 0.009197\\
\FeIII        & 1122.524 & 16.16~ -- 30.651& 0.07884\\
\FeII         & 1144.938 & ~7.870 -- 16.16~& 0.106\tablenotemark{d}\\
\FeII         & 1125.448 & ~7.870 -- 16.16~& 0.016\tablenotemark{d}\\
\HH\ 0-0 R(1) & 1108.634 & ~0~~~ -- 4.48\tablenotemark{b}&0.00117\tablenotemark{e}\\
\HH\ 4-0 R(2) & 1051.497 & ~0~~~ -- 4.48\tablenotemark{b}&0.0147\tablenotemark{e}\\
\HH\ 5-0 R(4) & 1044.546 & ~0~~~ -- 4.48\tablenotemark{b}&0.0162\tablenotemark{e}\\
\enddata
\tablenotetext{a}{Ionization potential ranges from \citet{Moore70} unless otherwise noted.} 
\tablenotetext{b}{\HH\ dissociation energies \citet{Spitzer78}.}
\tablenotetext{c}{$f$-values from \citet{Morton91} unless otherwise noted.}
\tablenotetext{d}{\citet{Howk00}.}
\tablenotetext{e}{\citet{Morton76}.}
\end{deluxetable}

\begin{deluxetable}{lrl}
\tabletypesize{\footnotesize}
\tablecolumns{3} 
\tablewidth{0pt} 
\tablecaption{Secondary Transitions Included in the Atlas}
\tablehead{\colhead{Ion}&\colhead{Wavelength}&\colhead{$f_\lambda$\tablenotemark{a}}}
\startdata
\HH\ 11-0 P(2) & 975.343 & 0.00663\\
\OI            & 976.448 & 0.003300\tablenotemark{b}\\
\HH\ 11-0 R(3) & 976.552 & 0.0121\\
\HH\ 11-0 P(3) & 978.216 & 0.00667\\
\HH\ 7-0 P(3) & 1019.506 & 0.0102\\
\HH\ 7-0 R(4) & 1020.768 & 0.0181\\
\HH\ 6-0 P(3) & 1031.191 & 0.0108\\
\HH\ 6-0 R(4) & 1032.356 & 0.0175\\
\HH\ 6-0 P(4) & 1035.184 & 0.0108\\
\HH\ 5-0 R(0) & 1036.546 & 0.0271\\
\CII*         & 1037.018 & 0.1230\tablenotemark{b}\\
\HH\ 5-0 R(1) & 1037.146 & 0.0185\\
\OVI          & 1037.617 & 0.0660\tablenotemark{b}\\
\HH\ 5-0 P(1) & 1038.156 & 0.00856\\
\HH\ 5-0 R(2) & 1038.690 & 0.0170\\
\HH\ 5-0 P(2) & 1040.367 & 0.00998\\
\HH\ 5-0 P(3) & 1043.498 & 0.0104\\
\HH\ 4-0 R(0) & 1049.366 & 0.0235\\
\HH\ 4-0 R(1) & 1049.958 & 0.0160\\
\HH\ 4-0 P(1) & 1051.031 & 0.00749\\
\HH\ 4-0 P(2) & 1053.281 & 0.00878\\
\HH\ 0-0 R(0) & 1108.128 & 0.00173\\
\HH\ 1-0 P(5) & 1109.313 & 0.00248\\
\HH\ 0-0 P(1) & 1110.063 & 0.000562\\
\HH\ 0-0 R(2) & 1110.120 & 0.00107\\
\HH\ 0-0 P(4) & 1120.247 & 0.000721\\
\HH\ 0-0 R(5) & 1120.399 & 0.000912\\
\FeII         & 1121.975 & 0.0202\tablenotemark{c}\\
\FeII         & 1127.089 & 0.0028\tablenotemark{c}\\
\FeII         & 1143.226 & 0.0177\tablenotemark{c}\\
\enddata
\tablenotetext{a}{$f$-values from \citet{Morton76} unless otherwise noted.}
\tablenotetext{b}{\citet{Morton91}.}
\tablenotetext{c}{\citet{Howk00}.}
\end{deluxetable}


\end{document}